\documentclass[twocolumn,prb,floatfix,superscriptaddress,citeautoscript,cite]{revtex4} %
\usepackage{graphicx}
\usepackage{epstopdf}
\usepackage{color}
\usepackage{amsmath}
\usepackage{soul}

\usepackage[caption=false,labelformat=empty,position=top,font=bf]{subfig}

\begin{document}

\title{Stacking-dependent topological magnons in bilayer CrI$_3$}

\author{M. Soenen}
\affiliation{Department of Physics \& NANOlab Center of Excellence, University of Antwerp, Groenenborgerlaan 171, B-2020 Antwerp, Belgium}

\author{C. Bacaksiz}
\affiliation{Department of Physics \& NANOlab Center of Excellence, University of Antwerp, Groenenborgerlaan 171, B-2020 Antwerp, Belgium}
\affiliation{Bremen Center for Computational Material Science (BCCMS), Bremen D-28359, Germany}
\affiliation{Computational Biotechnology, RWTH Aachen University, Worringerweg 3, 52074 Aachen, Germany}

\author{R. M. Menezes}
\affiliation{Department of Physics \& NANOlab Center of Excellence, University of Antwerp, Groenenborgerlaan 171, B-2020 Antwerp, Belgium}

\author{M. V. Milo\v{s}evi\'c}
\email{milorad.milosevic@uantwerpen.be}
\affiliation{Department of Physics \& NANOlab Center of Excellence, University of Antwerp, Groenenborgerlaan 171, B-2020 Antwerp, Belgium}

\begin{abstract}
Motivated by the potential of atomically-thin magnets towards tunable high-frequency magnonics, we detail the spin-wave dispersion of bilayer CrI$_3$. We demonstrate that the magnonic behavior of the bilayer strongly depends on its stacking configuration and the interlayer magnetic ordering, where a topological bandgap opens in the dispersion caused by the Dzyaloshinskii-Moriya and Kitaev interactions, classifying bilayer CrI$_3$ as a topological magnon insulator. We further reveal that both size and topology of the bandgap in a CrI$_3$ bilayer with an antiferromagnetic interlayer ordering are tunable by an external magnetic field.
\end{abstract}

\date{\today}

\maketitle

\section{Introduction}\label{sec:intro}
Emergent two-dimensional (2D) magnetic materials \cite{gibertini2019} provide an exciting platform to study collective spin excitations, i.e. magnons. CrI$_3$, the archetypal 2D van der Waals (vdW) ferromagnet\cite{huang2017}, has recently been suggested to host magnon modes in the highly sought-after terahertz (THz) regime\cite{jin2018,menezes2022,cenker2021}, showing promise for the development of faster and more energy-efficient data processing applications\cite{roadmap}. Moreover, due to the 2D nature of the material, its spin-wave properties are highly susceptible to tuning, e.g. by strain, buckling, defect-engineering, gating and/or vdW heterostructuring\cite{menezes2022}.

Recently, several bulk materials, including CrBr$_3$\cite{cai2021}, CrI$_3$\cite{chen2018,chen2021}, CrSiTe$_3$\cite{zhu2021} and CrGeTe$_3$\cite{zhu2021}, have been identified as \emph{topological magnon insulators} (TMIs), characterized by bulk magnon bands with a gap at the Dirac point, and topologically protected edge states. The magnonic bandgap is attributed to the anti-symmetric exchange interaction -- more often called the Dzyaloshinskii-Moriya interaction (DMI)\cite{dzialo1957,moriya1960} -- arising from the lack of inversion symmetry between next-nearest-neighboring (NNN) Cr atoms. In contrast, bulk CrCl$_3$\cite{chen2022,schneeloch2022}, where the DMI is weak, is classified as a \emph{magnon Dirac material} (MDM), characterized by a Dirac-point in the dispersion, showing a linear band crossing at the Brillouin zone edge. 

However, it remains an open question whether the topological features of aforementioned materials will persist down to the monolayer limit. Early theoretical work suggested that honeycomb ferromagnetic (FM) monolayers could be classified as either MDMs or TMIs depending on whether any NNN DMI is present in the material\cite{owerre2016,owerre2016-2,fransson2016,owerre2017-dm,pershoguba2018,kim2022}. Nonetheless, recent work identified Kitaev interactions as an alternative mechanism potentially able to open a topological bandgap in FM honeycomb materials\cite{aguilera2020,zhang2021}, suggesting that the absence of DMI is not the sole criterion for predicting the topological properties of such materials. Similarly, in magnetic honeycomb bilayers, a DMI-induced topological behavior of magnons is predicted\cite{kim2022,owerre2017-aa,owerre2016-ab,kondo2019}, including the formation of Dirac magnon nodal-line loops\cite{owerre2017-aa}, and the opening of a topological bandgap, which contributes to a magnon Hall- and a spin Nernst effect\cite{kim2022,owerre2016-ab,kondo2019}.

Several recent works\cite{costa2020,olsen2021,delugas-arxiv,gorni-arxiv} attempted to characterize the magnonics of monolayer CrI$_3$ using \emph{ab initio} calculations, and demonstrated the appearance of a small, possibly topological, bandgap caused by the spin-orbit coupling (SOC), suggesting that the material is a TMI. However, more work is required before full understanding of the magnonics in CrI$_3$ is achieved. In this work, we deploy a multi-scale approach, combining \emph{ab initio} calculations with numerical simulations based on a Heisenberg model and linear spin-wave theory, to characterize the magnonic properties of CrI$_3$ monolayers and bilayers, reveal the topological magnon modes present in these systems, and show that the (topological) magnonic properties of the bilayer are strongly affected by its stacking order and its interlayer magnetic ordering.

The article is organized as follows. In section \ref{sec:method}, we describe the computational methodology used in this work. We discuss the Heisenberg Hamiltonian that models the magnetic interactions in CrI$_3$, explain how the parameters that characterize this Hamiltonian will be derived from first-principles and, finally, sketch how the spin-wave dispersion is computed. Subsequently, in section \ref{sec:monolayer}, we apply this methodology to monolayer CrI$_3$, confirming the presence of a small topological bandgap with non-zero Chern numbers in the material's spin-wave dispersion. Afterwards, in section \ref{sec:bilayer}, we consider bilayer CrI$_3$ in three different stacking orders, each exhibiting significantly different magnonic behavior. Specifically, we investigate the AA-stacking and AB-stacking (rhombohedral) discussed in literature\cite{kim2022,owerre2017-aa,owerre2016-ab}, as well as the experimentally very relevant AB'-stacking (monoclinic) of which the spin-waves have - to the best of our knowledge -  not been theoretically investigated to date. We find that all three stacking versions of bilayer CrI$_3$ exhibit either FM or antiferromagnetic (AFM) interlayer ordering, with intralayer ferromagnetism. In case of a FM interlayer ordering, we observe a bandgap in the spin-wave dispersion with stacking-dependent topological properties. We attribute the origin of the gap to a combination of DMI and Kitaev interactions that are modulated by the stacking order. Furthermore, we show a significant influence of the interlayer magnetic ordering on the resulting magnonic behavior. Specifically, the topological nature of the bands becomes trivial in AFM-ordered bilayers. Additionally, we show that magnonic dispersion of AFM-ordered bilayers is susceptible to tuning by an external magnetic field, lifting the degeneracy between branches, which decreases the size of the magnonic bandgap and leads to a non-trivial topology of the bands in the AB'-stacking, or introduces nodal-line loops in the AA-stacking case. Finally, section \ref{sec:conclu} summarizes our findings and gives an outlook on some future challenges and opportunities within the field.

\section{Computational methodology}\label{sec:method}
We model the magnetic interactions of the system under study using a Heisenberg spin Hamiltonian of the following form:
\begin{equation}\label{eq:ham}
 \hat{\mathcal{H}}=\frac{1}{2}\sum_{i , j} \mathbf{\hat{S}}_i \mathcal{J}_{ij} \mathbf{\hat{S}}_j + \sum_i \mathbf{\hat{S}}_i \mathcal{A}_{ii}\mathbf{\hat{S}}_i + \mu_\mathrm{B}\sum_i\mathbf{B}\cdot g_i \mathbf{\hat{S}}_i,
\end{equation} 
in which the spins are three-dimensional (3D) vectors $\mathbf{\hat{S}}_i = (\hat{S}_{i}^{x},\hat{S}_{i}^{y},\hat{S}_{i}^{z})$ expressed in Cartesian coordinates. The first- and second term of this Hamiltonian respectively describe the exchange interaction and the single ion anisotropy (SIA), which are characterized by the $3 \times 3$ matrices $\mathcal{J}_{ij}$ and $\mathcal{A}_{ii}$. The DMI is characterized by a vector $\mathbf{D}_{ij}$ with components that can be calculated from the off-diagonal elements of the exchange matrix as $D^x_{ij} = \frac{1}{2}(\mathcal{J}^{yz}_{ij}-\mathcal{J}^{zy}_{ij})$, $D^y_{ij} = \frac{1}{2}(\mathcal{J}^{zx}_{ij}-\mathcal{J}^{xz}_{ij})$ and $D^z_{ij} = \frac{1}{2}(\mathcal{J}^{xy}_{ij}-\mathcal{J}^{yx}_{ij})$\cite{sabani2020,xiang2013}. Notice that $\mathbf{D}_{ij} = \nu_{ij}|\mathbf{D}_{ij}|$ with $\nu_{ij} = -\nu_{ji} = \pm 1$, where the sign of the latter depends on the hopping direction of the considered spin pair. The exchange term is now written as:
\begin{align}\label{eq:ex}
    \hat{\mathcal{H}}_\mathrm{ex} = \frac{1}{2}\sum_{i , j} \Big[\sum_{\alpha'} J_{ij}^{\alpha'} \hat{S}_i^{\alpha'}  \hat{S}_j^{\alpha'} + \mathbf{D}_{ij} (\mathbf{\hat{S}}_i \times \mathbf{\hat{S}}_j)\Big],
\end{align}
with $\alpha' = \{\alpha,\beta,\gamma\}$ the local eigenbases that diagonalize the symmetric part of the exchange matrices. To consider the exchange anisotropy, we define the Kitaev constant as $K_{ij} = J_{ij}^{\gamma} - J_{ij}$ with $J_{ij} = (J_{ij}^\alpha + J_{ij}^\beta)/2$ the isotropic exchange constant\cite{xu2018}, leading to the following expression for the exchange Hamiltonian: 
\begin{align}
    \hat{\mathcal{H}}_\mathrm{ex} = \frac{1}{2}\sum_{i , j} \left[J_{ij} \mathbf{\hat{S}}_i \cdot \mathbf{\hat{S}}_j + K_{ij} \hat{S}_i^{\gamma}  \hat{S}_j^{\gamma} + \mathbf{D}_{ij} (\mathbf{\hat{S}}_i \times \mathbf{\hat{S}}_j)\right]. \nonumber
\end{align}
The symmetric SIA-matrix $\mathcal{A}_{ii}$ accounts for the interaction of the magnetic orbitals with the surrounding crystal field and contributes to the magnetic anisotropy of the material. In crystals with a 3-, 4-, or 6-fold rotational symmetry around the out-of-plane axis, most elements of the matrix are redundant and the SIA can be characterized by a single parameter $\mathcal{A}_{ii}^{zz}$ instead of the full SIA-matrix\cite{sabani2020,xiang2013}. The last term of equation (\ref{eq:ham}) accounts for the Zeeman interaction when applying an external magnetic field $\mathbf{B}$, where $g_i\approx 2$ is the $g$-factor, and $\mu_\mathrm{B}$ is the Bohr magneton. In CrI$_3$, the magnetic dipole-dipole interaction is expected to be small in comparison to its out-of-plane magnetic anisotropy and will, therefore, not be included in the Heisenberg Hamiltonian\cite{menezes2022}. Finally, also notice that CrI$_3$ has a magnetic moment of $\mu = 3\mu_\mathrm{B}$ per chromium atom and, thus, a spin of S = 3/2.

To obtain the elements of the exchange- and SIA matrices, we use the four-state energy mapping (4SM) methodology\cite{sabani2020,xiang2013} in which we calculate the energies of several spin configurations of the system from first principles using density functional theory (DFT), and map these energies on their corresponding Heisenberg Hamiltonians, setting up a system of equations from which the magnetic parameters can be derived. The implementation of the needed DFT calculations is thoroughly discussed in section S.I of the supplementary material\cite{sup_mat}.

The spin-wave dispersion relations are calculated numerically using the open-source code \emph{spinW}\cite{spinw}, in which we have implemented our Heisenberg Hamiltonian. This code is based on linear spin-wave theory, which is a good approximation assuming spin fluctuations are small. This condition is comfortably satisfied at low temperatures, significantly below the critical temperature (Curie or N\'eel) of the long-range magnetic order at hand. Numerical diagonalization of the Heisenberg Hamiltonian in reciprocal space yields the spin-wave dispersion.

\section{Monolayer}\label{sec:monolayer}
\subsection{Crystal structure and magnetic parameters}
\begin{figure*}[tp!]
\includegraphics[width=0.95\linewidth]{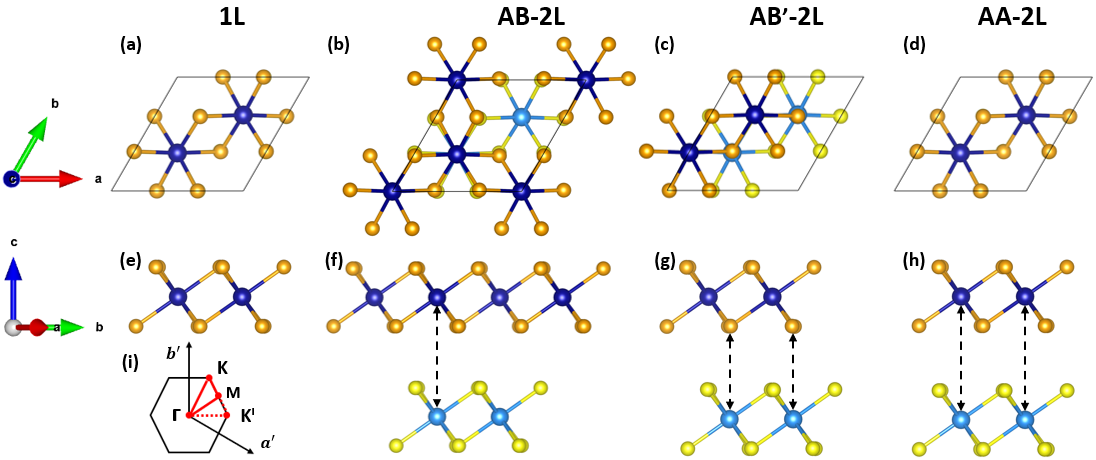}
\caption{\label{fig:structure} \textbf{Crystal structure of monolayer and bilayer CrI$_\mathbf{3}$.} Top view (a-d) and side view (e-h) of monolayer (1L) CrI$_3$ (a,e), and bilayer (2L) CrI$_3$ with an AB-stacking (b,f), AB'-stacking (c,g) and AA-stacking (d,h). For the sake of clarity, atoms of the same type are assigned a different color in the top and the bottom layer. In the bottom (top) layer, the chromium and iodine atoms are depicted with blue (dark blue) and yellow (orange) spheres respectively. The unit cell is marked with a solid black line. All crystal structures were plotted using VESTA\cite{VESTA}. Panel (i) depicts the corresponding first Brillouin zone and high-symmetry points for 2D systems with a hexagonal lattice.}
\end{figure*}

The crystal structure of monolayer CrI$_3$ is depicted in Figure~\ref{fig:structure}(a,e). Monolayer CrI$_3$ comprises one honeycomb layer of chromium atoms sandwiched between two layers of iodine atoms, where each chromium atom is octahedrally coordinated with six iodine atoms, and each iodine atom connects two chromium atoms through an $\approx 90^{\circ}$ Cr-I-Cr bond. After structural relaxation using DFT, we find a in-plane lattice constant of $a$ = 6.919 \AA{}. 

To characterize the magnetic interactions in CrI$_3$, we perform a 4SM analysis in order to obtain the elements of the exchange and SIA matrices. In Table \ref{tab:parameters_mono}, we report the average nearest-neighbor (NN) and next-nearest-neighbor (NNN) intralayer exchange, Kitaev and DMI parameters for monolayer CrI$_3$. A full summary of the exchange parameters of all the individual pairs can be found in section S.IV of the supplementary material\cite{sup_mat}. Notice that third nearest-neighbor (3NN) and higher order exchange terms are not taken into account as their influence on the spin-wave dispersion is negligible\cite{sup_mat}.
\begin{table}[bp!]
\caption{\label{tab:parameters_mono} \textbf{Magnetic parameters for monolayer CrI$_\mathbf{3}$.} Summary of the most important magnetic parameters in monolayer CrI$_3$, including the exchange- and Kitaev constants $J_{ij}$ and $K_{ij}$, and the size of the DMI-vectors $|\mathbf{D}_{ij}|$.}
\begin{tabular*}{\columnwidth}{c @{\extracolsep{\fill}} ccccc}
\hline\hline
$J_{\mathrm{NN}}$ & $K_{\mathrm{NN}}$ & $|\mathbf{D}_{\mathrm{NN}}|$ & $J_{\mathrm{NNN}}$ & $K_{\mathrm{NNN}}$ & $|\mathbf{D}_{\mathrm{NNN}}|$ \\
(meV) & (meV) & (meV) & (meV) & (meV) & (meV) \\ \hline
-4.35 & 1.49 & 0.00 & -0.74 & 0.17 & 0.06 \\
\hline\hline 
\end{tabular*}
\end{table}

From the calculated parameters, it becomes clear that both the NN and the NNN exchange interactions are anisotropic and FM, with the NN one delivering the dominant contribution. In agreement with literature\cite{lado2017,bacaksiz2021}, we find that the material's out-of-plane magnetic anisotropy originates mainly from the NN exchange anisotropy, with a smaller contribution of $\langle \mathcal{A}^{zz}_{ii}\rangle$ = -0.08 meV due to the SIA. The SIA is characterized by a single parameter owing to the material's three-fold rotational symmetry. The NN interactions deliver no net contribution to the DMI since the inversion symmetry of the material is upheld. However, this symmetry is not present between NNN sites, resulting in a small yet non-zero DMI. Notice that, in CrI$_3$, the DMI, the Kitaev interaction and the SIA all originate from the large SOC arising due to the heavy iodine ligands\cite{xu2018,lado2017,bacaksiz2021}.

\subsection{Spin-wave dispersion}
\begin{figure*}[tp!]
\centering
\subfloat[\raggedright(a)]{\includegraphics[width=.32\textwidth]{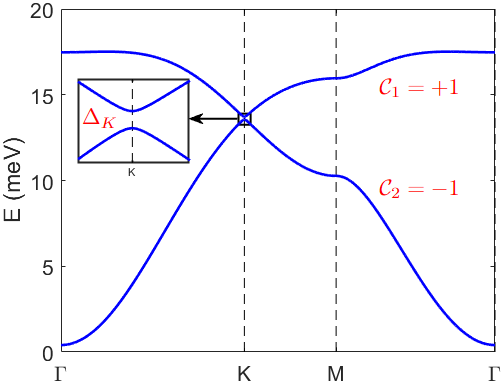}}
\hfill
\subfloat[\raggedright(b)]{\includegraphics[width=.16\textwidth]{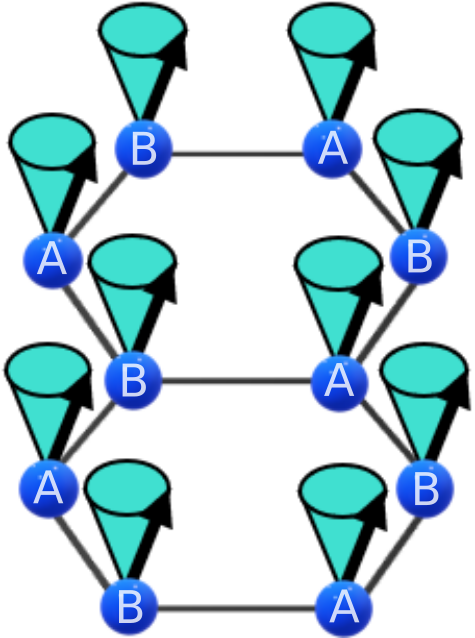}}
\hfill
\subfloat[\raggedright(c)]{\includegraphics[width=.16\textwidth]{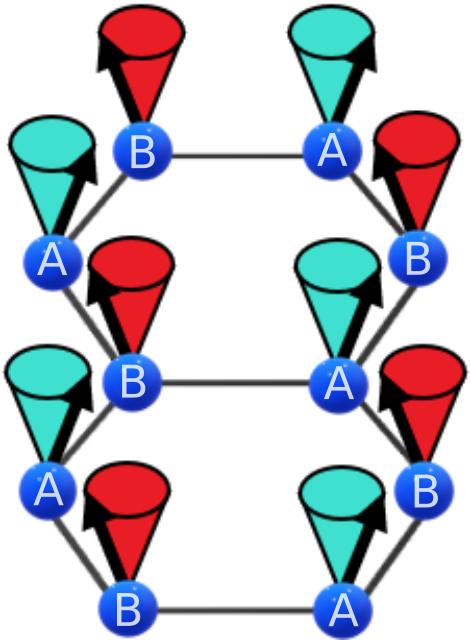}}
\hfill
\subfloat[\raggedright(d)]{\includegraphics[width=.13\textwidth]{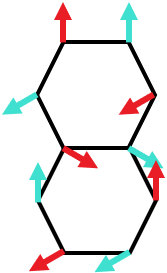}}
\hfill
\subfloat[\raggedright(e)]{\includegraphics[width=.11\textwidth]{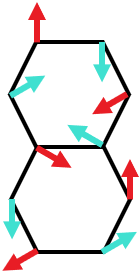}}
\caption{\label{fig:disp_mono} \textbf{Spin-wave dispersion of monolayer CrI$_\mathbf{3}$.} (a) Spin-wave dispersion along the high-symmetry directions of the first Brillouin zone. A small Dirac gap of $\Delta_K = 0.15$ meV is present at the K-point. Corresponding Chern numbers are indicated for each band. (b) and (c) display the spin-wave modes at the $\Gamma$-point for the lower- and higher branch respectively. (d) and (e) display a schematic top-view of the spin-wave modes at the K-point for the lower- and higher branch respectively.}
\end{figure*}

Figure \ref{fig:disp_mono}(a) depicts the spin-wave dispersion of monolayer CrI$_3$ along the high-symmetry directions of the first Brillouin zone. Two distinct branches are present, as is expected for a unit cell containing two magnetic atoms. At the $\Gamma$-point, the dispersion shows a Goldstone gap due to the magnetic anisotropy of the material. The gap has a size of $\Delta_\Gamma = 0.43$ meV and is an essential prerequisite for the existence of long-range magnetic order in 2D systems at finite temperatures\cite{lado2017}. The latter can be seen by considering the total number of magnons excited at temperature $T$, which is given by:
\begin{align}\label{eq:n_mag}
    N = \int \frac{\mathcal{D}(\omega_\mathbf{k})}{e^{\hbar \omega_\mathbf{k}/k_BT}-1}d\omega_\mathbf{k},
\end{align}
with $\mathcal{D}(\omega_\mathbf{k})$ the magnon density of states, which is constant in 2D, $k_B$ the Boltzmann constant, and $\omega_\mathbf{k}$ the spin-wave frequencies. When the dispersion is gapless, i.e. in the absence of magnetic anisotropy, this integral will diverge for $\omega_\mathbf{k} = 0$, preventing long-range 2D magnetic order at non-zero temperature in accordance with the Mermin-Wagner theorem\cite{mermin-wagner}.

The lower energy `acoustic' branch displays quadratic behavior near the $\Gamma$-point and is associated with an in-phase precession of the spins [see Figure 2(b)], while the higher energy `optical' branch is associated with an out-of-phase precession of the spins [see Figure 2(c)]. The two branches meet at the K-point where they are separated by a small bandgap of $\Delta_K = 0.15$ meV. At the K'-point we find a gap of the same size, since the sublattice symmetry is upheld. The origin of this Dirac gap is partially attributed to the NNN DMI and partially to the Kitaev interactions. The determined bandgap is smaller but comparable in size to Ref. [\onlinecite{olsen2021}] and Ref. [\onlinecite{gorni-arxiv}], who report bandgaps of respectively 0.30 meV and 0.47 meV. Differences might be attributed to the use of different methodologies and \emph{ab initio} parameters. However, these theoretically predicted bandgaps all differ significantly from the experimentally observed one of 2.8 meV in bulk CrI$_3$ \cite{chen2021} (Currently, there are no measurements for monolayer CrI$_3$ available yet). Note that some recent work\cite{costa2020} found a bandgap of 2 meV, which is significantly closer to the experimental value. The origin of this discrepancy between theory and experiment is still open to debate, however, magnon-phonon interactions\cite{delugas-arxiv} -- whose effect is not captured in our model -- are suggested to be responsible for the enhancement of the bandgap. Despite the quantitative differences, our results show a good qualitative agreement with the earlier mentioned works.

At the K-point ($\lambda = 3a/2$), the spins will precess at 120$^\circ$ angles to each other, as is shown in Figure 2(d,e) for respectively the lower- and higher branch. If one would only consider a purely isotropic NN exchange, these two states would be energetically degenerate resulting in a Dirac point. However, introducing Kitaev interactions and/or a NNN exchange term which includes a non-zero DMI component, lifts the mutual degeneracy between the modes resulting in a bandgap.

More specifically, the magnonic bandgap is attributed to the NN Kitaev interactions and the out-of-plane component of the NNN DMI. The size of the latter which is intrinsically present in CrI$_3$ is rather small ($|D^{z}_{ij}| = 0.03$ meV), also resulting in a small bandgap. However, external tuning that breaks the inversion symmetry, e.g. the presence of a substrate, electric gating or (non-uniform) strain, might induce additional DMI that could potentially increase the size of the bandgap. In fact, by (artificially) increasing the DMI in our simulations, we verified that the bandgap can be `tuned'. As shown in section S.V of the supplementary material\cite{sup_mat}, the bandgap scales almost linearly with the NNN DMI when all the other parameters are kept constant. Tuning the magnonic bandgap in 2D materials under external stimuli poses an interesting direction for future research, as the size of the bandgap can influence other material properties such as the magnon Hall conductivity. However, note that increasing the DMI may lead to non-collinear magnetization textures, e.g. spin cycloids or magnetic skyrmions, which will fundamentally change the magnonic behavior in the material\cite{ma2015,weber2022}. Moreover, when the DMI is set to zero in our calculations, the bandgap does not fully vanish, suggesting that there is a second mechanism at work, which we identify to be the Kitaev interaction between NN spins. In section S.V of the supplementary material\cite{sup_mat}, we show that one can tune the bandgap by artificially changing $K_{ij}$, and present a phase diagram how both the size and topology of the bandgap vary as a function of $D^{z}_{\mathrm{NNN}}$ and $K_{\mathrm{NN}}$. However, one should bear in mind that varying the strength of the Kitaev interaction also influences the overall shape of the dispersion, whereas changing the DMI mainly influences the dispersion around the K-point.  

\begin{table*}[tp!]
\caption{\label{tab:parameters_bilayer} \textbf{Structural and magnetic parameters for bilayer CrI$_\mathbf{3}$.} Summary of the most important stucutural and magnetic parameters in bilayer CrI$_3$, including the lattice constant $a$ and interlayer distance $d$, the average exchange- and Kitaev constants $\langle J_{ij}\rangle$ and $\langle K_{ij}\rangle$, the average size of the DMI-vectors $\langle |\mathbf{D}_{ij}|\rangle$, the DFT energy difference between the bilayer with an AFM and a FM interlayer ordering, and the average SIA.}
\begin{tabular*}{\textwidth}{l|@{\extracolsep{\fill}} cccccccccc}
\hline\hline
& $a$ & $d$ & $\langle J_{\mathrm{NN}}\rangle$ & $\langle K_{\mathrm{NN}}\rangle$ & $\langle|\mathbf{D}_{\mathrm{NN}}|\rangle$ & $\langle J_{\mathrm{NNN}}\rangle$ & $\langle K_{\mathrm{NNN}}\rangle$ & $\langle|\mathbf{D}_{\mathrm{NNN}}|\rangle$ & $E_{\mathrm{AFM}}-E_{\mathrm{FM}}$ & $\langle \mathcal{A}^{zz}_{ii}\rangle$ \\
& (\AA{}) & (\AA{}) & (meV) & (meV) & (meV) & (meV) & (meV) & (meV) & (meV) & (meV) \\ \hline
AB  & 6.915 & 3.400 & -4.49 & 1.45 & 0.07 & -0.62 & 0.13 & 0.03 &  12.13 & -0.07 \\
AB' & 6.914 & 3.430 & -4.49 & 1.45 & 0.07 & -0.64 & 0.15 & 0.02 & -0.06 & -0.07 \\
AA  & 6.908 & 3.505 & -4.42 & 1.44 & 0.07 & -0.65 & 0.15 & 0.03 &  0.84 & -0.08 \\
\hline\hline 
\end{tabular*}
\end{table*}

\subsection{Topology}
Non-trivial band topology arises only in systems where non-zero Chern numbers predict the existence of edge states. The Chern number is a topological invariant with an integer value that is defined for the $n$th band as:
\begin{align}\label{eq:chern}
    \mathcal{C}_n = \frac{1}{2\pi i}\int_\mathrm{BZ} \Omega_{n\mathbf{k}}\ d^2k,
\end{align}
in which the Berry curvature can be calculated as
\begin{align}
    \Omega_{n\mathbf{k}} = i \sum_{n^{\prime} \neq n} \frac{\langle n\left|\partial_\mathbf{k} \right. \hat{\mathcal{H}}_\mathbf{k} \left| n^{\prime} \right. \rangle \langle n^{\prime}\left|\partial_\mathbf{k} \right. \hat{\mathcal{H}}_\mathbf{k} \left| n\right. \rangle}{\left(\lambda_{n\mathbf{k}}-\lambda_{n^{\prime}\mathbf{k}}\right)^2},
\end{align}
with $\lambda_{n\mathbf{k}}$ and $|n\rangle$ respectively the eigenvalues and eigenvectors of the Heisenberg Hamiltonian $\hat{\mathcal{H}}_\mathbf{k}$ in reciprocal space. For systems that are gapless or show a trivial bandgap, the Chern numbers vanish. In this work, we calculate Chern numbers according to the link-variable method detailed in Ref. [\onlinecite{fukui2005}] for a discretized Brillouin zone. Applying this approach to the magnonic dispersion of monolayer CrI$_3$, we find Chern numbers of $\mathcal{C}_n = \pm 1$ for respectively the upper and lower band, as shown in Figure~\ref{fig:disp_mono}, classifying the material as a TMI with a non-trivial topological bandgap. We attribute the origin of the topology to the breaking of time-reversal symmetry due to the spontaneous magnetization of CrI$_3$\cite{mcclarty2022}. Thus, we can conclude that the topological nature of the bands persists in monolayer CrI$_3$, be it with a significantly smaller bandgap compared to bulk CrI$_3$\cite{chen2021}.

\section{Bilayer}\label{sec:bilayer}
\subsection{Crystal structure and magnetic parameters}
Bilayer CrI$_3$ can be constructed by stacking two monolayers on top of each other in a commensurate manner. The three different stacking orders that we consider in this work are shown in Figure \ref{fig:structure}. In analogy to Sivadas \textit{et al.}\cite{sivadas2018}, we refer to those stacking orders as AB (rhombohedral), AB' (monoclinic), and AA. The former two stackings correspond to respectively the low-temperature and the high-temperature phases of CrI$_3$\cite{mcguire2015}. In the AB-stacking, the layers are stacked in such a way to place the chromium atoms in one layer above the void in the chromium honeycomb of the adjacent layer, analogously to a Bernal-stacked graphene bilayer [Figure~\ref{fig:structure}(b,f)]. The AB-stacking can be transformed to an AB'-stacking by sliding one of the layers by a third of the lattice vector along the zigzag direction [Figure~\ref{fig:structure}(c,g)]. Alternatively, by sliding one of the AB-stacked layers by a third of the lattice vector along the armchair direction, we obtain an AA-stacked bilayer in which each atom in the top layer is placed exactly above its bottom layer counterpart [Figure~\ref{fig:structure}(d,h)]. 

As shown in Table~\ref{tab:parameters_bilayer}, the different stacking orders show relatively similar lattice constants and interlayer distances. However, changes in interatomic distances and (super-)superexchange bonding angles result in a different interlayer magnetic coupling, such that the AB and AA stackings prefer a FM ordering between the layers while the AB'-stacking slightly favors an AFM one. The latter is indicated in Table~\ref{tab:parameters_bilayer} by the DFT energy difference between AFM and FM phases. We found similar results in simulations using the Landau-Lifshitz-Gilbert (LLG) equation in \emph{Spirit}\cite{spirit} and the Metropolis Monte Carlo algorithm in \emph{spinW}\cite{spinw}. In agreement with earlier theoretical work\cite{sivadas2018,jang2019,jiang2019,soriano2019}, we find that the overall ground state of the system is a FM-ordered AB-stacked bilayer. At first sight, the calculated FM ground state is at odds with experiment\cite{huang2017,thiel2019,song2019,li2019,chen2019}, where an AFM state is observed for even layered CrI$_3$, however, recent work by Thiel \emph{et al.}\cite{thiel2019} suggests that the FM interlayer ordering might be the ground state for freestanding CrI$_3$, in agreement with most DFT studies, while encapsulation of CrI$_3$ causes a structural and magnetic phase transition to a state with an AFM interlayer ordering. In section S.II of the supplementary material\cite{sup_mat}, we discuss the stacking-dependence of the interlayer ordering in more detail. 

In Table \ref{tab:parameters_bilayer}, we also summarize the predominant magnetic parameters for the CrI$_3$ bilayers calculated with the 4SM method. A full overview of all the calculated parameters for each specific pair can be found in section S.IV of the supplementary material\cite{sup_mat}. For all stacking orders, the NN intralayer exchange interaction is anisotropic and strongly FM. This anisotropy, together with the SIA, causes the spins to prefer an out-of-plane orientation. Due to the rotational symmetry in the AB and AA-stackings, the SIA matrix is reduced to only one parameter $\mathcal{A}^{zz}_{ii}$. In the AB'-stacked bilayer, this symmetry is absent requiring a calculation of the full SIA-matrix, however, $\mathcal{A}^{zz}_{ii}$ will still be the dominant parameter, as most of the other matrix elements are very small or vanish. 

To quantify the interlayer coupling, we calculated the interlayer NN and NNN exchange matrices for all stackings. For the AB'-stacking, we also calculate the 3NN interlayer exchange, for the other stackings this contribution is negligible as is demonstrated in section S.IV of the supplementary material\cite{sup_mat}. In the AB- and AA-stacked bilayers, all NN and NNN interlayer exchange interactions are FM. However, the exchange parameters for the AB-stacking are significantly stronger than for the AA-stacking, resulting in a stronger preference for a FM ordering. In contrast, for the AB'-stacked bilayer, there is a competition between the NN exchange which is FM and the NNN and 3NN exchange interactions which are AFM. Overall, this results in a weak AFM interlayer ordering, which is in agreement with earlier theoretical and experimental studies\cite{huang2017,sivadas2018,jang2019,jiang2019,soriano2019,thiel2019,song2019,li2019,chen2019,jiang2021-switch}. 

\begin{figure*}[tp!]
\centering
\hspace*{\fill}
\subfloat[(a) \textbf{AB}]{\includegraphics[width=.42\textwidth]{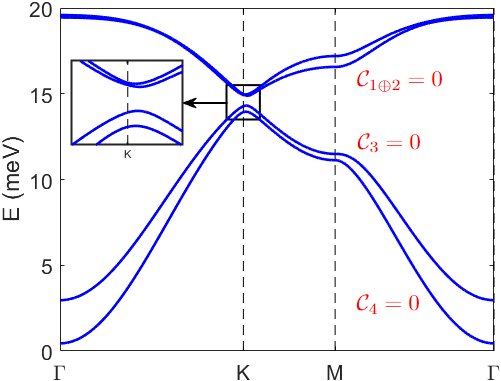}}
\hfill
\subfloat[(b) \textbf{AB'}]{\includegraphics[width=.42\textwidth]{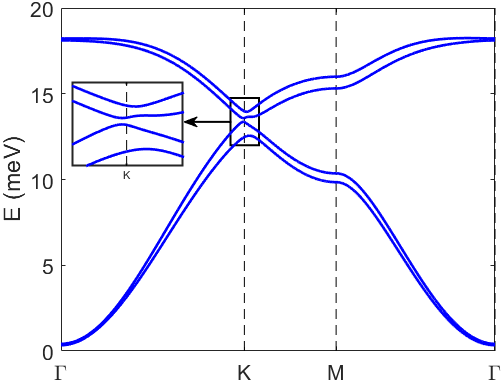}} \hspace*{\fill} \\
\hspace*{\fill}
\subfloat[(c) \textbf{AA}]{\includegraphics[width=.42\textwidth]{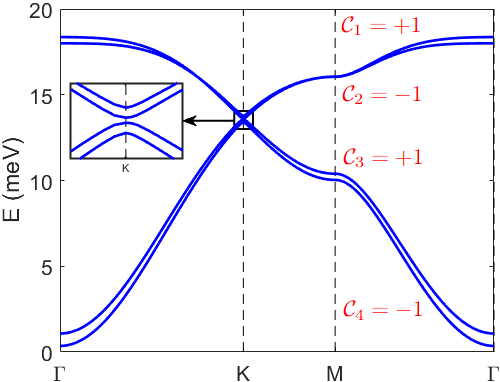}}
\hfill
\subfloat[(d)]{\includegraphics[width=.42\textwidth]{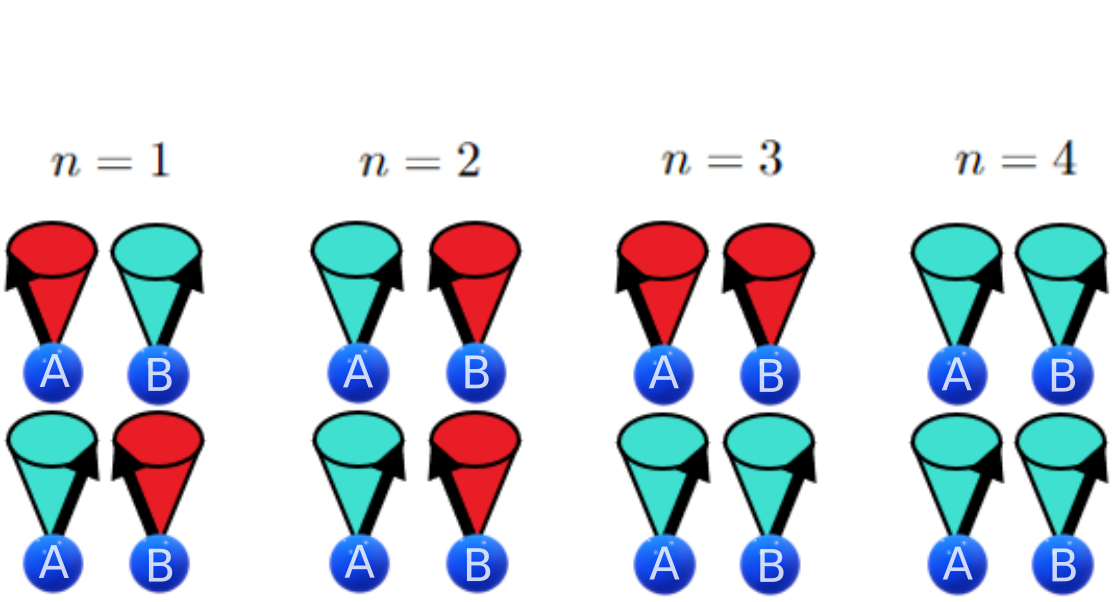}} \hspace*{\fill}
\caption{\label{fig:disp_fm} \textbf{Spin-wave dispersion for bilayer CrI$_\mathbf{3}$ with a FM interlayer ordering in different stacking configurations.} (a,b,c) Spin-wave dispersion along the high-symmetry directions of the first Brillouin zone for respectively the AB, AB', and AA-stacked bilayers with a FM interlayer ordering. In the AB and AA-stackings, a direct magnonic bandgap opens close to the K-point. In the AB'-stacking, there is an indirect band crossing. Corresponding Chern numbers are indicated for single bands, and composite Chern numbers for degenerate bands. (d) schematically displays the corresponding spin-wave modes at the $\Gamma$-point for each band.}
\end{figure*}

Interestingly, the sublattice symmetry is broken in the AB- and AB' stackings, leading to a difference in out-of-plane exchange interactions $\Delta J^{zz} = |J^{zz}_\mathrm{A} - J^{zz}_\mathrm{B}|$ between sublattices A and B of 0.92 meV for the AB-stacking and 0.04 meV for the AB'-stacking. Note that $J^{zz}_\mathrm{A}$ and $J^{zz}_\mathrm{B}$ are the sum of the out-of-plane exchange components of all interacting spin pairs in a unit cell. The difference $\Delta J^{zz}$ is substantial for the AB-stacking because one sublattice has six stronger interlayer NNN interactions while the other sublattice has one weaker interlayer NN coupling and only three interlayer NNN interactions. For the AA-stacking, there is no exchange difference since the sublattice symmetry is preserved. 

The intralayer Kitaev constants in the bilayers are similar in size compared to the monolayer. For the AB- and AB'-stacking, the NN Kitaev interaction is anisotropic, leading to different values for each bond, which is attributed to symmetry breaking due to the stacking. Since the NN Kitaev interaction is much stronger than the NNN and the interlayer ones, it's the only contribution having a significant influence on the spin-wave dispersion.

Unlike the monolayer system, the NN intralayer DMI is now non-zero, and originates from the inversion symmetry breaking due to stacking. Similarly to the monolayer case, a non-zero NNN DMI arises. In all stackings, the interlayer DMI will be very small or completely absent, having a limited influence on the dispersion.

\begin{figure*}[tp!]
\centering
\hspace*{\fill}
\subfloat[(a) \textbf{AB}]{\includegraphics[width=.42\textwidth]{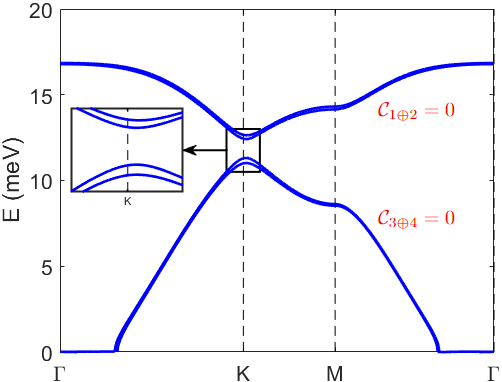}}
\hfill
\subfloat[(b) \textbf{AB'}]{\includegraphics[width=.42\textwidth]{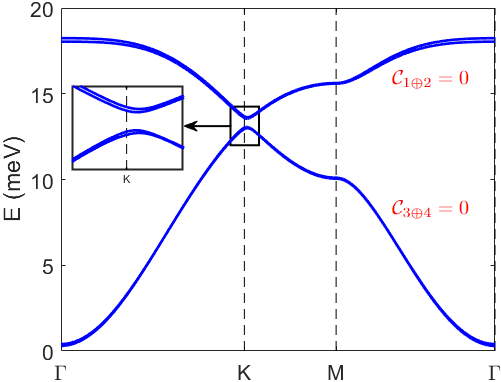}} \hspace*{\fill} \\
\hspace*{\fill}
\subfloat[(c) \textbf{AA}]{\includegraphics[width=.42\textwidth]{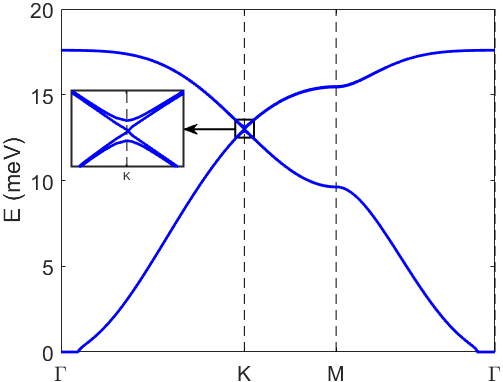}}
\hfill
\subfloat[(d)]{\includegraphics[width=.42\textwidth]{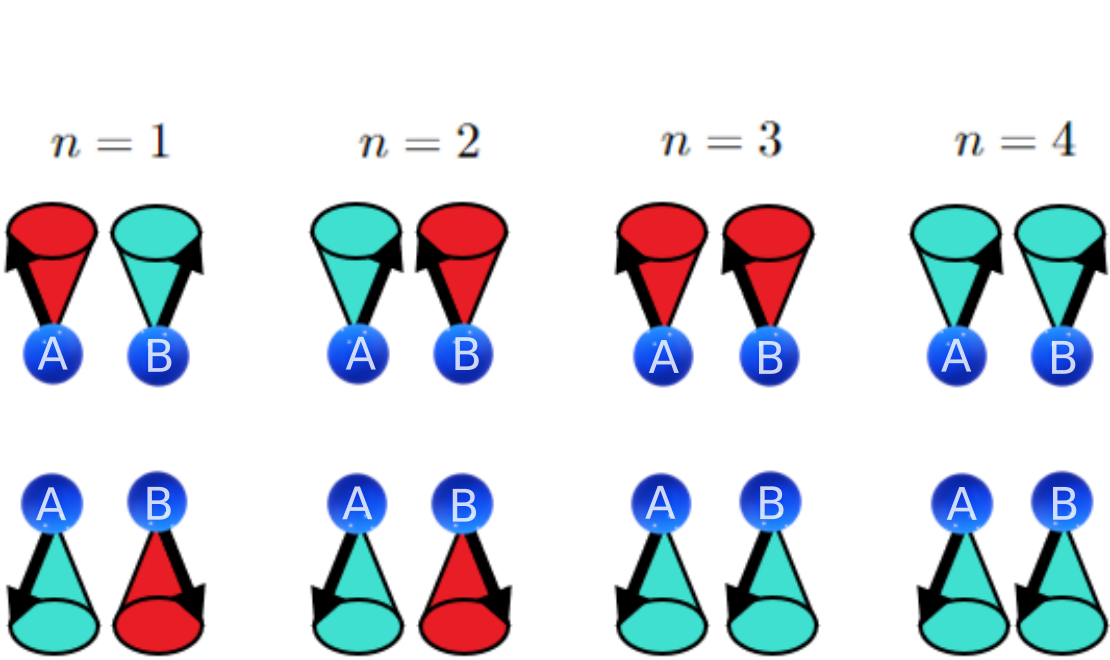}} \hspace*{\fill}
\caption{\label{fig:disp_afm} \textbf{Spin-wave dispersion for bilayer CrI$_\mathbf{3}$ with an AFM interlayer ordering in different stacking configurations.} (a,b,c) Spin-wave dispersion along the high-symmetry directions of the first Brillouin zone for respectively the AB, AB' and AA-stacked bilayers with an AFM interlayer ordering. In the AB and AB'-stackings a direct magnonic bandgap opens close to the K-point, meanwhile in the AA-stacked bilayer we observe a Dirac point. Corresponding composite Chern numbers are indicated for the bands and are all equal to zero. (d) schematically displays the corresponding spin-wave modes at the $\Gamma$-point for each band.}
\end{figure*}

\subsection{Spin-wave dispersion of bilayers with FM interlayer order}
Using the magnetic parameters calculated with the 4SM method, we compute the spin-wave dispersion for the three stacking orders considered in this work. In Figure~\ref{fig:disp_fm}, we display the results for bilayer CrI$_3$ with different stackings, all with the FM interlayer ordering. In a bilayer, the unit cell contains four magnetic atoms, leading to four branches in the dispersion, two `acoustic' and two `optical' ones. The corresponding spin-wave modes at the $\Gamma$-point of each branch are indicated in Figure \ref{fig:disp_fm}(d). The energy difference between these different modes is proportional to the strength of the interlayer coupling, hence the large separation for the AB-stacking. 

Each stacking shows a Goldstone gap at the $\Gamma$-point, signaling that FM order is stable in each of them at finite temperatures. The gaps have sizes of $\Delta_\Gamma$ = 0.44~meV, $\Delta_\Gamma$ = 0.34~meV and $\Delta_\Gamma$ = 0.35~meV for respectively the AB, AB', and AA-stackings. In the AB and AA-stackings, we observe direct magnonic bandgaps close to the K-point of $\Delta_{K^*} = 0.37$~meV, and $\Delta_K = 0.07$~meV for respectively the AB, and AA-stackings. Similarly to the monolayer, we attribute the origin of these gaps to a combination of NNN DMI and NN Kitaev interactions. In the AB'-stacking, there is an indirect band crossing, as is often seen in semi-metals, and thus, no bandgap. Notice that the bandgaps don't occur exactly at $\mathbf{K}$ but are shifted to $\mathbf{K^*}$ due to the breaking of in-plane honeycomb symmetry caused by the interlayer coupling\cite{gorni-arxiv}. For the AB-stacking we see a shift of $\mathbf{K} - \mathbf{K^*} = (-0.005,0.021)\frac{2\pi}{a}$. In the AB'-stacking, we see shifts of $(0.037,-0.019)\frac{2\pi}{a}$ and $(0.008,-0.005)\frac{2\pi}{a}$ for respectively the n=2 and n=3 bands. In the AA-stacking, there is no shift. The first Brillouin zone contains two inequivalent high-symmetry points K and K' [Figure~\ref{fig:structure}(i)]. In the AB and AB'-stackings, where the sublattice symmetry is broken, we see a different behavior of the dispersion at each point. In the former, a bandgap of only 0.05~meV opens close to the K'-point (compared to $\Delta_{K^*} = 0.37$~meV). In both the AB and AB'-stackings, we see shifts from the the K'-point to the K'$^*$-point of rougly the same size but in the opposite direction. In the AA-stacked bilayer, the sublattice symmetry is preserved, resulting in exactly the same dispersion at both the K- and K'-points. 

\subsection{Spin-wave dispersion of bilayers with AFM interlayer order}
By comparing the dispersion of the bilayers with FM interlayer ordering with the dispersions of the bilayers with AFM interlayer ordering [Figure~\ref{fig:disp_afm}], it becomes clear that there is a strong dependence of the magnonic properties of CrI$_3$ on the interlayer ordering. First and foremost, notice that for the AA and AB stackings, there is a region close to the $\Gamma$-point where the acoustic branches are zeroed. Consequently, there is no gap at the $\Gamma$-point and the integral in equation \eqref{eq:n_mag} will diverge, signaling that AFM order is unstable in these stackings at non-zero temperatures. However, in the AB'-stacking, there is a gap of $\Delta_\Gamma = 0.30$~meV, meaning that AFM order is stable in the monoclinic phase, which is in agreement with experimental observations\cite{huang2017,mcguire2015,thiel2019,song2019,li2019,chen2019}. Further, also notice that, in contrast to the FM-ordered bilayers, we see a degeneracy of the two acoustic branches and the two optical branches. Only at the K-point there are notable energy differences between the bands. 

The dispersions of the bilayers with AFM interlayer order are characterized by bandgaps of respectively $\Delta_{K^*}$ = 1.04~meV and $\Delta_{K'^*}$ = 0.97~meV for the AB-stacking, with shifts of ($\mathbf{K} - \mathbf{K^*}) = -(\mathbf{K'} - \mathbf{K'^*}) = (-0.005,0.009)\frac{2\pi}{a}$, and $\Delta_{K^*}$ = $\Delta_{K'^*}$ = 0.11~meV for the AB'-stacking, with shifts of ($\mathbf{K} - \mathbf{K^*}) = -(\mathbf{K'} - \mathbf{K'^*}) = (-0.005,-0.008)\frac{2\pi}{a}$. At the K and K'-points in the AA-stacking case, there is no bandgap, but instead one finds a Dirac cone combined with two anti-crossing branches. 

\subsection{Topology}
When two or more bands are degenerate, crossing or touching, it is no longer possible to assign individual Chern numbers to each band. Instead, we define a composite Chern number $\mathcal{C}_{n \oplus n'}$, jointly shared by the degenerate bands, and calculated as detailed in Ref.~[\onlinecite{zhao2020}]. 

As shown in Figure \ref{fig:disp_fm}, there is a strong dependence of the Chern number on the stacking configuration in the FM-ordered bilayers. Although we are expecting a non-trivial topology of the bandgaps in the FM bilayers, caused by the breaking of time-reversal symmetry due to the spontaneous magnetization, only the AA-stacking shows non-zero Chern numbers. Thus, the AA-stacked CrI$_3$ bilayer can be classified as a TMI. In the AB'-stacking there is no bandgap and, hence, the Chern number is undefined and the bands are not topological. In the AB-stacking, all Chern numbers are equal to zero, meaning that the bandgap is of trivial nature. We attribute the lack of topology in the latter stacking to the exchange difference $\Delta J^{zz}$, caused by the breaking of sublattice symmetry, which is very large for the AB-stacking. In section S.V of the supplementary material\cite{sup_mat}, we show that by artificially reducing $\Delta J^{zz}$ in our simulations, which also decreases the size of the bandgap at the K-point and the K'-point, we can induce a topological phase transition to a state with non-zero Chern numbers of $\mathcal{C}_{1 \oplus 2} = +1$, $\mathcal{C}_3 = -1$ and $\mathcal{C}_{4} = 0$, which confirms the influence that sublattice symmetry can have on the topology of magnonic bands\cite{kim2022}.

In bilayers with AFM interlayer order, the bands are two-by-two degenerate, meaning that one can only define composite Chern numbers. In the case of AA-stacking, there is no bandgap and, thus, the Chern number is undefined and the bands display no topological behavior. The composite Chern numbers for the other two stackings turn out to be zero for all considered bands, which can be related to the conservation of effective time-reversal symmetry in AFM materials, as the layers are time-reversed copies of each other\cite{mcclarty2022}. However, in the next section, we will show that breaking this symmetry by an applied magnetic field leads to emergent topological states with non-zero Chern numbers.

\subsection{Effect of an external magnetic field}
In this section, we explore whether the magnonic dispersion and band topology of bilayer CrI$_3$ can be tuned by applying an out-of-plane external magnetic field. 

\begin{figure}[bp!]
\centering
\includegraphics[width=.49\columnwidth]{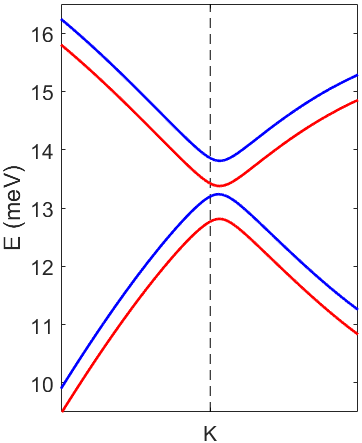}
\includegraphics[width=.49\columnwidth]{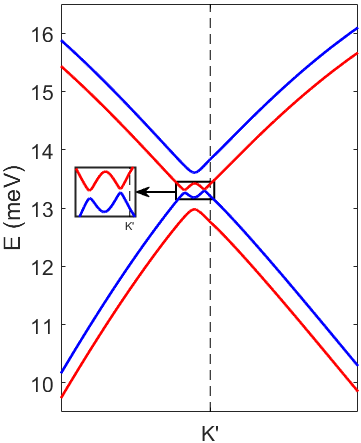}
\caption{\label{fig:disp_abp_field} \textbf{Spin-wave dispersion for an AB'-stacked CrI$_\mathbf{3}$ bilayer with an AFM interlayer ordering under the influence of an external magnetic field.} Left and right panels compare the dispersion near the K-point and K'-point respectively, for an applied magnetic field of $B = 1.9$~T. Under influence of the magnetic field, blue bands have shifted up and red bands down in energy.}
\end{figure}

In the case of a monolayer or the bilayers with FM interlayer order, there is only a trivial effect due to an applied magnetic field. Namely, the whole dispersion will uniformly shift up or down depending on the orientation of the applied field with respect to the magnetization. Notice that, a very large oppositely oriented field can flip the magnetization, which changes the sign of the Chern number of each band, meaning that the propagation direction of the magnonic edge states reverses. 

In contrast, for bilayers with AFM interlayer order, an external magnetic field will lift the degeneracy between branches, shifting two branches up and two branches down in energy, and leading to additional interesting features in the dispersion. Similar band shifts were observed by Cenker \textit{et al.}\cite{cenker2021}, who reported the splitting of degenerate Raman peaks in the optical branches of an AFM-ordered monoclinic CrI$_3$ bilayer, after applying an external magnetic field. However, note that applying a magnetic field to AFM-ordered bilayers should be done carefully, as the interlayer magnetic state will switch to the FM one for sufficiently strong fields. Here, we calculate the spin-wave dispersion for the different stacking scenarios under sufficiently small applied field, where AFM interlayer order is safely stable (see section S.III of the supplementary material\cite{sup_mat}). Especially the AFM-AB phase is very sensitive, and changes to a FM interlayer order even for very weak applied field - hence is excluded from our calculations in this section. 

\begin{figure}[tp!]
\centering
\subfloat[\qquad B = 0.0 T]{\includegraphics[width=.49\columnwidth]{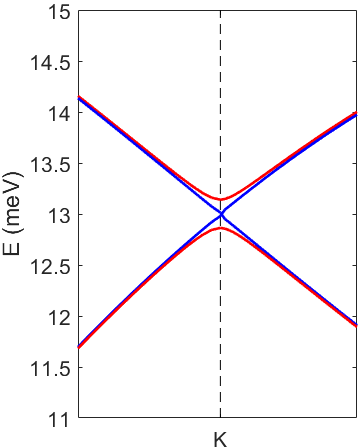}}
\subfloat[\qquad B = 1.6 T]{\includegraphics[width=.49\columnwidth]{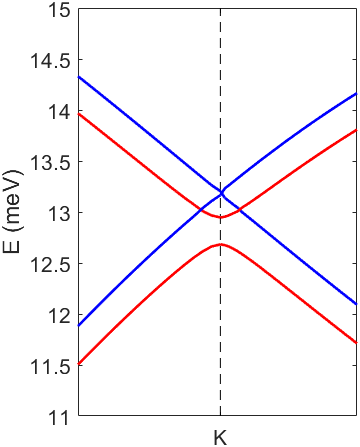}}
\caption{\label{fig:disp_aa_field} \textbf{Spin-wave dispersion for an AA-stacked CrI$_\mathbf{3}$ bilayer with an AFM interlayer ordering under the influence of an external magnetic field.} Left and right panels compare the dispersion at the K-point for applied fields of respectively $B = 0$~T and $B = 1.6$~T. Under influence of the magnetic field, blue bands have shifted up and red bands down in energy, forming magnon nodal-lines at the crossing points of the red and blue curves.}
\end{figure}

In the case of an AB' stacking, the size of the bandgap will decrease after applying the magnetic field, reaching a minimum 0.01~meV for a field of $B = 1.9$~T, as shown in Figure~\ref{fig:disp_abp_field}. On the right panel, one sees that, as the bands approach each other, they do not entirely touch or cross - instead we observe band inversion combined with a small bandgap. Band inversion is an effect often also present in electronic topological insulators\cite{zhu2012}, and is typically caused by the SOC. For fields applied in the opposite direction, we see analogous behavior, as now the two other bands are shifted upwards and the previous two downwards. For fields larger than $B = 1.9$~T, the AFM interlayer ordering changes to a FM one (see section S.III of the supplementary material\cite{sup_mat}).

In Figure~\ref{fig:disp_aa_field}, we show the influence of an external magnetic field with a magnitude of $B = 1.6$~T on the dispersion of an AFM-ordered AA-stacked CrI$_3$ bilayer. Applying the field shifts the Dirac node upwards or downwards depending on the polarity of the field, which leads to the formation of a closed Dirac magnon nodal-line loop at the crossover point of the red and blue bands in Figure~\ref{fig:disp_aa_field}. The latter is a closed one-dimensional loop around the K-point where two bands cross, exactly analogous to the nodal-lines described for Dirac semimetals\cite{wehling2014}. Decreasing the field leads to a smaller shift of the branches, resulting in nodal-line loops with a smaller radius. For fields larger than $1.6$~T, the AFM interlayer ordering changes to a FM one (see section S.III of the supplementary material\cite{sup_mat}).

As mentioned earlier, in the absence of applied field, all AFM bands show composite Chern numbers equal to zero, meaning that the bandgaps have a trivial topology. Interestingly, after applying the magnetic field on the AB'-stacked bilayer, non-zero Chern numbers emerge as $\mathcal{C}_{1,4} = +1$ and $\mathcal{C}_{2,3} = -1$. In other words, by applying a magnetic field, which breaks the effective time-reversal symmetry of the material, a topological phase transition can be induced. In contrast, for the AA stacking, the (composite) Chern numbers remain undefined after applying the magnetic field, as the Dirac cone stays present. 

\section{Conclusions}\label{sec:conclu}
We characterized the magnonic dispersion for intrinsically ferromagnetic monolayer and bilayer CrI$_3$ using linear spin-wave theory combined with a Heisenberg model parameterized from first principles. We showed that the monolayer is characterized by a small Dirac-gap in the spin-wave dispersion, sourced to a specific combination of next-nearest-neighbor (NNN) DMI and nearest-neighbor Kitaev interactions. Non-zero Chern numbers are associated with the bands, indicating the topological nature of the bandgap, and suggesting that monolayer CrI$_3$ is a topological magnon insulator (TMI). In bilayer CrI$_3$, still with dominantly ferromagnetic intralayer interactions, we demonstrated a dependence of the dispersion on the geometric stacking order and the interlayer magnetic ordering, opening a bandgap for the AB stacking (for both FM and AFM interlayer order), the AB' stacking (only AFM), and the AA stacking (only FM), meanwhile the FM-ordered AB' stacking shows an indirect band crossing, and the AFM-ordered AA stacking exhibits a Dirac point. Similarly to the monolayer case, we identified the DMI and Kitaev interactions as the leading causes behind the opening of the bandgap, both being modulated by the stacking order. The latter contradicts earlier work on bulk CrI$_3$ which claimed that only the NNN DMI and, thus, not the Kitaev interaction, lies at the origin of the Dirac gap\cite{chen2021}. Interestingly, we found that the Chern number, and consequently the magnonic band topology, depends on the stacking configuration and the interlayer magnetic order, vanishing for all studied cases except in the FM-ordered AA bilayer. Thus, depending on the stacking order and the interlayer magnetic order, bilayer CrI$_3$ is classified as either a \textit{topological} magnon insulator, a \textit{trivial} magnon insulator, or a \textit{magnon Dirac material}. Finally, we showed that the dispersion of the bilayers with AFM interlayer order can be tuned by an external out-of-plane magnetic field, changing both size and topology of the bandgap for the AB'-stacked bilayer, and introducing closed nodal-line loops in the dispersion of the AA bilayer. 

The here demonstrated presence of tunable bandgaps of possibly topological nature in bilayer CrI$_3$ recommend it as a TMI that can serve as a platform to investigate tunable magnon Hall- and spin Nernst effects in 2D. Our results could be verified experimentally by investigating the thermal magnon Hall effect in monolayer and bilayer CrI$_3$. Both the DMI and Kitaev interactions originate from the spin-orbit coupling, which is relatively strong in CrI$_3$ and, thus, lies at the origin of the topological bandgap. If one wants to achieve a gapless spin-wave dispersion, we suggest looking at 2D magnets with a weaker SOC, e.g. CrBr$_3$ or CrCl$_3$, which are good candidates to host a Dirac point in the monolayer limit. In order to further tailor the magnonic bandgap, one can induce and tune the DMI in CrI$_3$, or other 2D magnets, by external stimuli such as gating, (non-uniform) strain, heterostructuring, etc. Furthermore, our work demonstrates that stacking vdW monolayers, be it in regular bilayers, or in future work multilayers and (moir\'e) heterostructures, poses a viable route to achieve broadly tunable magnonic properties in 2D materials and van der Waals homo- and heterostructures.   

\begin{acknowledgments}
The authors thank D. \v{S}abani, B. Jorissen and M. Shafiei for useful discussions. This work was supported by the Research Foundation-Flanders (FWO-Vlaanderen) and the Special Research Funds of the University of Antwerp (BOF-UA). The computational resources used in this work were provided by the VSC (Flemish Supercomputer Center), funded by Research Foundation-Flanders (FWO) and the Flemish Government -- department EWI.
\end{acknowledgments}

\end{document}


\title{Stacking-dependent topological magnons in bilayer CrI$_3$\\ SUPPLEMENTARY MATERIAL}

\author{M. Soenen}
\affiliation{Department of Physics \& NANOlab Center of Excellence, University of Antwerp, Groenenborgerlaan 171, B-2020 Antwerp, Belgium}

\author{C. Bacaksiz}
\affiliation{Department of Physics \& NANOlab Center of Excellence, University of Antwerp, Groenenborgerlaan 171, B-2020 Antwerp, Belgium}
\affiliation{Bremen Center for Computational Material Science (BCCMS), Bremen D-28359, Germany}
\affiliation{Computational Biotechnology, RWTH Aachen University, Worringerweg 3, 52074 Aachen, Germany}

\author{R. M. Menezes}
\affiliation{Department of Physics \& NANOlab Center of Excellence, University of Antwerp, Groenenborgerlaan 171, B-2020 Antwerp, Belgium}

\author{M. V. Milo\v{s}evi\'c}
\email{milorad.milosevic@uantwerpen.be}
\affiliation{Department of Physics \& NANOlab Center of Excellence, University of Antwerp, Groenenborgerlaan 171, B-2020 Antwerp, Belgium}

\date{\today}

\maketitle

\section{DFT calculations}\label{supsec:dft}
The DFT calculations are performed within the Vienna ab initio simulation package (VASP)\cite{vasp1,vasp2,vasp3} using a generalized gradient approximation (GGA) functional and the projector augmented wave (PAW) method\cite{paw}. We opt for the Perdew-Burke-Ernzerhof (PBE)\cite{pbe} exchange-correlation functional in combination with the D2 method of Grimme\cite{grimme} to account for a vdW correction term. A term due to the SOC\cite{soc} is added ad hoc to the DFT Hamiltonian where necessary. 

\emph{Vacuum distance} --- In VASP, periodic boundary conditions are implemented automatically. Hence, during the calculations, we choose an out-of-plane unit cell distance of $c = 15$ \AA{} for the monolayer and $c = 26$ \AA{} for the bilayer ensuring a large enough vacuum distance between the materials and their periodic images, effectively minimizing the interaction between them, and leading to a converged total energy and pressure in these systems.

\emph{Supercell size} --- Due to the periodic boundary conditions, we need to use 3 $\times$ 3 $\times$ 1 supercells during the 4SM calculations in order to calculate the NNN and 3NN interactions. This is the smallest possible supercell yielding converged NNN and 3NN exchange parameters, for smaller supercells these values will be overestimated due to interaction with periodic images. However, the use of supercells increases the computational cost of our calculations, hence, we will try to limit this cost by choosing the smallest possible energy cutoff and number of k-points for which we deem the system sufficiently converged.

\begin{figure}[!b]
  \centering
  \subfloat[][]{\includegraphics[width=.45\textwidth]{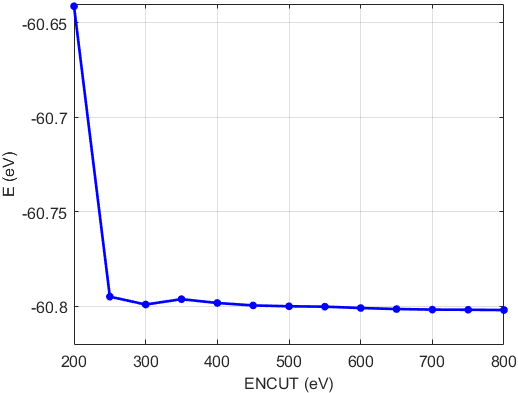}}
  \subfloat[][]{\includegraphics[width=.48\textwidth]{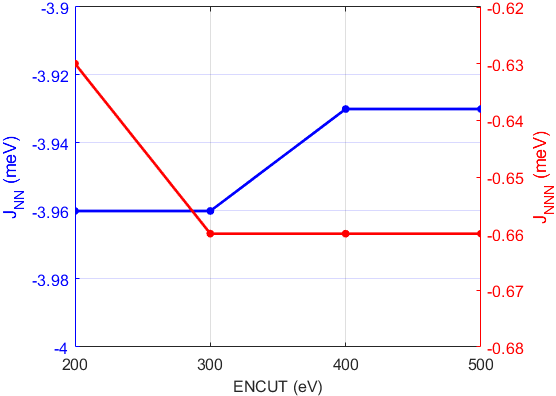}}
  \caption{\textbf{Convergence test for the ENCUT parameter used in the DFT calculations.} (a) Total energy of a CrI$_3$ monolayer as a function of the energy cutoff. (b) The J$_{\mathrm{NN}}^{zz}$ and the J$_{\mathrm{NNN}}^{zz}$ exchange parameters of monolayer CrI$_3$ for increasing ENCUT values.}
  \label{fig:encut}
\end{figure}

\emph{Plane-wave energy cutoff} --- In Figure \ref{fig:encut}(a), we show how the total energy of the system converges as a function of the energy cutoff, which is implemented by the ENCUT parameter in VASP. For the structural relaxation of our systems, we use a plane-wave energy cutoff of 700 eV. To limit the computational cost during the 4SM calculations, we lower the cutoff to 300 eV, for which the system is converged within 0.003 eV. As shown in Figure \ref{fig:encut}(b), using a smaller energy cutoff has very limited influence on the NN and NNN exchange parameters, resulting in deviations smaller than 0.03 meV, having a negligible quantitative and no qualitative impact on our results. 

\begin{figure}[!t]
  \centering
  \subfloat[][]{\includegraphics[width=.46\textwidth]{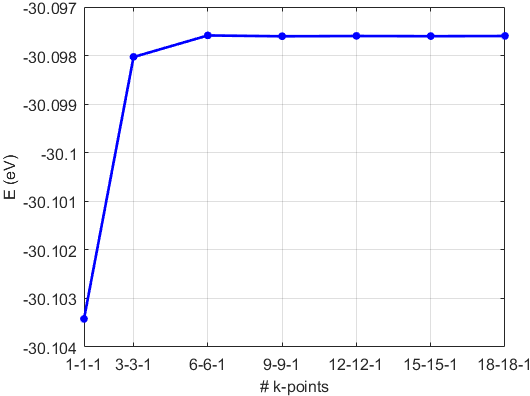}}
  \subfloat[][]{\includegraphics[width=.48\textwidth]{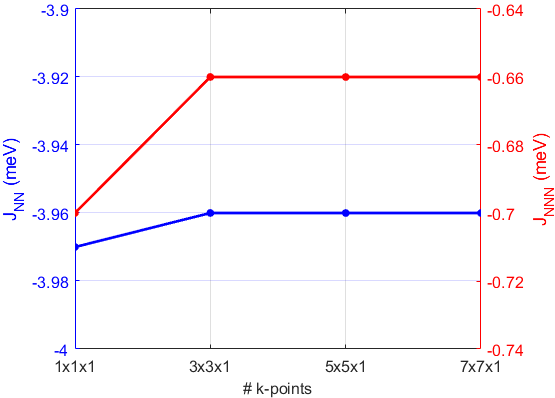}}
  \caption{\textbf{Convergence test for the number of k-points used in the DFT calculations.} (a) Total energy of a CrI$_3$ monolayer as a function of the k-point grid. (b) The J$_{\mathrm{NN}}^{zz}$ and the J$_{\mathrm{NNN}}^{zz}$ exchange parameters of monolayer CrI$_3$ for an increasing number of k-points.}
  \label{fig:kpoints}
\end{figure}

 \emph{Brillouin zone sampling} --- For the Brillouin zone integration, we use a Gaussian smearing of 0.01 eV. During the structural relaxation of our systems, we use a $15 \times 15 \times 1$ grid for the k-point sampling. The use of supercells in the 4SM calculations justifies the use of a smaller k-point grid. To limit the computational cost, we opt for a $3 \times 3 \times 1$ k-point grid, for which the total energy is converged within 0.002 eV as shown in Figure \ref{fig:kpoints}(a). Furthermore, in Figure \ref{fig:kpoints}(b), we show that additionally enlarging the k-point grid has zero impact on the calculated exchange parameters.  

\begin{figure}[!ht]
  \centering
  \includegraphics[scale=0.59]{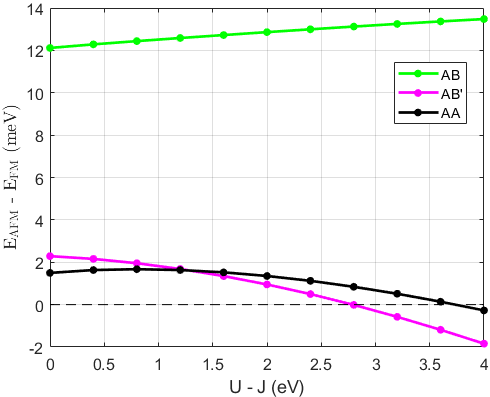}
  \caption{\textbf{Variation of the magnetic ground-state in bilayer CrI$_3$ as a function of the effective U-parameter used in the DFT calculations.} The total energy difference between the AFM and FM phases of the three considered stacking orders of bilayer CrI$_3$ as a function of the on-site Coulomb interaction U - J (with J fixed to 1.1 eV).}
  \label{fig:sup_dft}
\end{figure}

\emph{U-parameter} --- Since the system contains localized (strongly correlated) d-electrons, we implement the GGA+U method in the rotationally invariant form proposed by Dudarev et al.\cite{dudarev}, i.e. we add an on-site Coulomb interaction of U$^{\mathrm{eff}}$ = U - J = 2.8 eV to the d-orbitals of the chromium atoms. To confirm that the chosen U$^{\mathrm{eff}}$ gives the desired qualitative description of the magnetic properties of CrI$_3$, we plot the energy difference between the AFM and FM phases as a function U - J (with J = 1.1 eV) for all stacking orders [Fig. \ref{fig:sup_dft}]. For the AB-stacked bilayer (green curve), we find a strong preference for a FM interlayer ordering for any value of U - J. Similarly, for the AA-stacking (black curve), we find a (smaller) preference for a FM interlayer ordering, except for very big U - J values. For the AB'-phase (magneta curve), however, we find a FM interlayer ordering for small U - J values which transitions to an AFM ordering for increasing U - J. Similarily to Jiang et al.\cite{jiang2019}, we choose values of U = 3.9 eV and J = 1.1 eV resulting in an effective parameter of U - J = 2.8 eV which slightly favors an AFM-coupled AB'-stacking and a FM-coupled AB- and AA-stackings. This agrees with our value of U = 2.8037 eV that we calculated with a linear response method for monolayer CrI$_3$.

\newpage
\section{Stacking dependence of the interlayer coupling} \label{supsec:stacking}
\begin{figure}[!ht]
  \centering
  \hspace*{\fill}
  \subfloat[][]{\includegraphics[width=.45\textwidth]{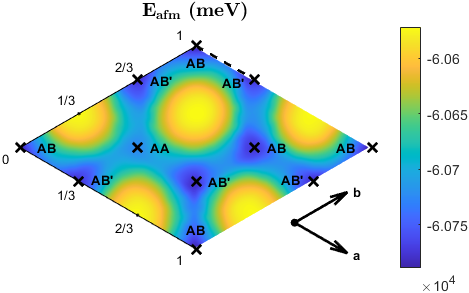}}\hfill
  \subfloat[][]{\includegraphics[width=.45\textwidth]{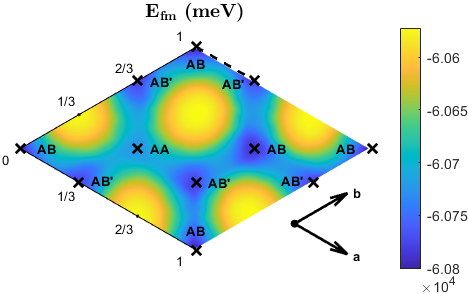}} \hspace*{\fill} \\
  \hspace*{\fill}
  \subfloat[][]{\includegraphics[width=.45\textwidth]{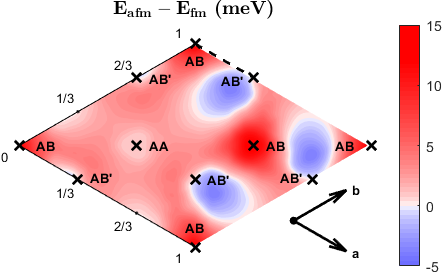}}\qquad \hfill
  \subfloat[][]{\includegraphics[width=.36\textwidth]{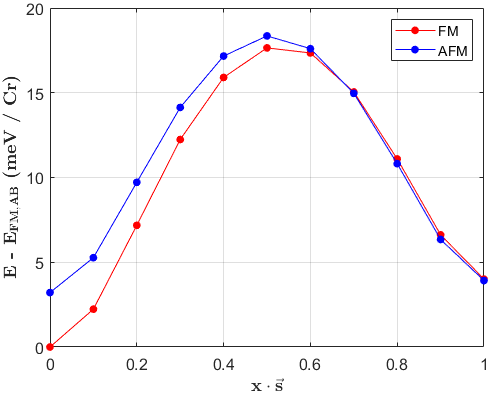}} \qquad \qquad \hspace*{\fill}
  \caption{\textbf{Total energy of bilayer CrI$\mathbf{_3}$ as a function of a lateral shift with respect to an AB-stacking.} The total energy is depicted as a function of a lateral shift of the top layer along basis vectors \textbf{a} and \textbf{b}, for a bilayer with AFM (a) and FM (b) interlayer ordering respectively. Minima in the energy correspond to the AB, AB' and AA-stacking orders. The energy difference between the AFM and FM phases is shown in (c). Panel (d) shows the energy per Cr atom, with respect to the ground-state (AB,FM), along the transition pathway that is marked with a black dashed line in panels (a-c).}
  \label{fig:sup_stacking}
\end{figure}

To map out the stable phases of bilayer CrI$_3$, we plot its energy as the top layer is shifted laterally [Fig. \ref{fig:sup_stacking} (a,b)]. The structures are relaxed in the out-of-plane direction to find the optimal interlayer distance. For both an AFM [Fig. \ref{fig:sup_stacking} (a)] and a FM [Fig. \ref{fig:sup_stacking} (b)] ordered bilayer, we find similar energy profiles with clear minima located at the naturally occurring rhombohedral (AB) and monoclinic (AB') phases, and  a local minimum at the AA stacking order, hence our choice for these stackings during this work. To determine the preferential interlayer magnetic order for each stacking, we plot the energy difference between the AFM and FM phases in Figure \ref{fig:sup_stacking} (c). Stacking orders that prefer an AFM configuration show up in blue, phases with a preference for a FM ordering show up in red. In the groundstate, the system will take on an AB-stacking combined with a FM interlayer coupling. The AA-stacked bilayer also prefers a FM interlayer coupling. Although the energy difference is very small, the AB'-stacked bilayer prefers an AFM interlayer coupling. This small energy difference is better illustrated in Figure \ref{fig:sup_stacking} (d) which depicts an intersection of the AFM and FM energy profiles along the pathway marked with a black dashed line in Figure \ref{fig:sup_stacking} (a-c). The results reported here are in good agreement with earlier work by Jiang et al. \cite{jiang2019} and Sivadas et al. \cite{sivadas2018}. All results presented in Figure \ref{fig:sup_stacking} were obtained from first-principles using DFT calculations implemented in VASP. 

\newpage
\section{Magnetic field effect on the magnetic order}\label{supsec:bfield}
To further investigate the dynamic stability of the different phases of bilayer CrI$_3$, both with and without the presence of an external magnetic field, we perform atomistic spin dynamics (ASD) simulations, by numerically solving the Landau-Lifshitz-Gilbert (LLG) equation:
\begin{align}\label{eq:llg}
\frac{\partial \hat{\mathbf{S}}_{i}}{\partial t}=-\frac{\gamma}{\left(1+\alpha^{2}\right) \mu}\left[\hat{\mathbf{S}}_{i} \times \mathbf{B}_{i}^{\mathrm{eff}}+\alpha \hat{\mathbf{S}}_{i} \times\left(\hat{\mathbf{S}}_{i} \times \mathbf{B}_{i}^{\mathrm{eff}}\right)\right],   
\end{align}
in which $\gamma$ is the gyromagnetic ratio, $\alpha = 0.001$ the damping parameter, $\mu = 3\mu_\mathrm{B}$ the magnetic moment per Cr atom, and $\mathbf{B}_{i}^{\mathrm{eff}}=-\partial \hat{\mathcal{H}} / \partial \hat{\mathbf{S}}_{i}$ the effective field. The latter is the resulting field due to all the magnetic interactions considered in the Heisenberg Hamiltonian (see main text). The simulations are performed at T = 0 K on a 50 $\times$ 50 supercell for a duration of $10^5$ iterations. The ASD simulations are implemented in the \emph{Spirit}\cite{spirit} software package which is adapted to accommodate the Hamiltonian considered in this work. In the remainder of this section we first investigate the magnetic stability without an applied magnetic field. Afterwards, we discuss the influence of an external out-of-plane magnetic field on both the FM and AFM states of monolayer and bilayer CrI$_3$.

\begin{figure}[!b]
  \centering
  \includegraphics[width=.75\textwidth]{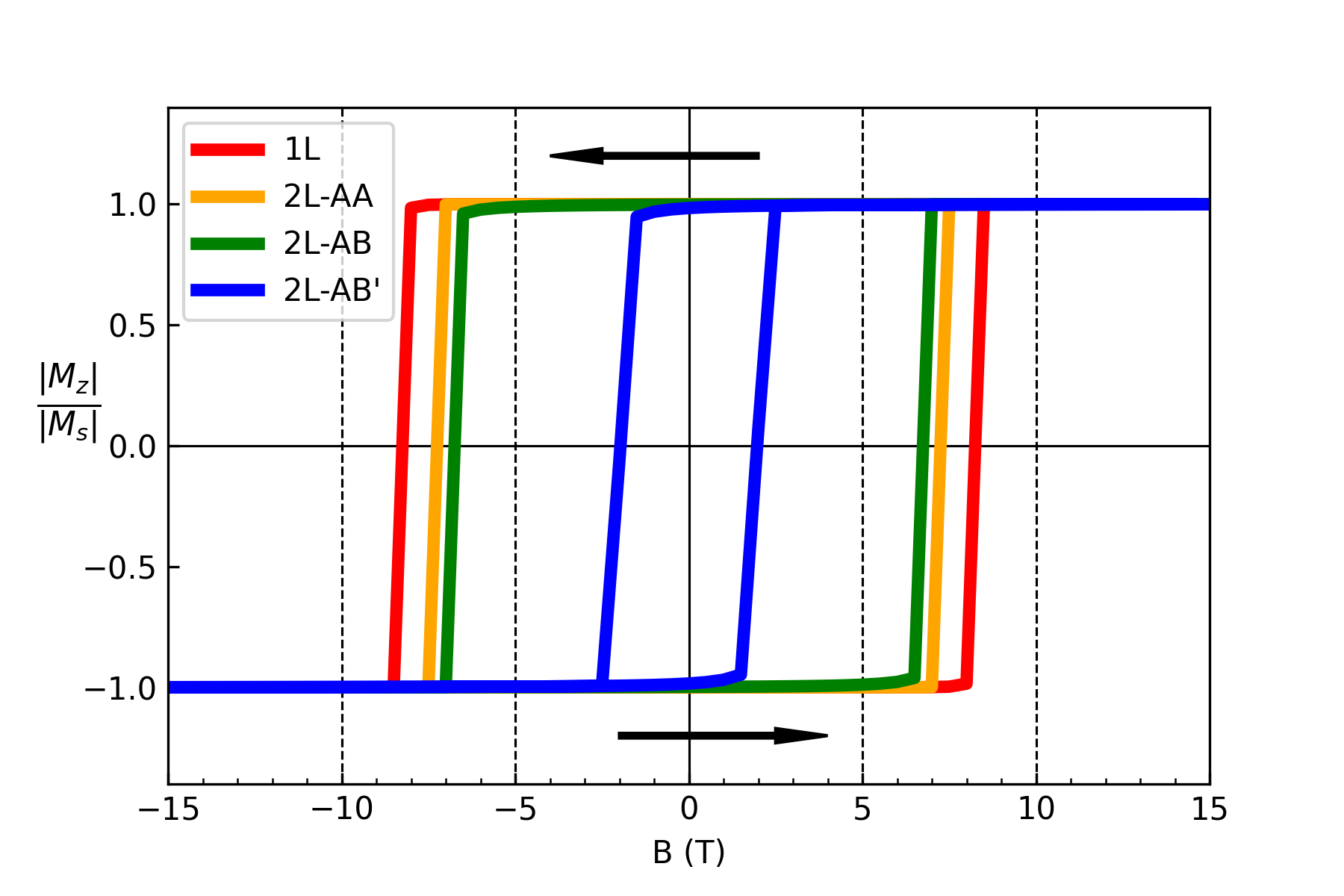}
  \caption{\textbf{Hysteresis curves for monolayer and bilayer CrI$_3$.} Out-of-plane magnetization of monolayer (1L) and bilayer (2L) CrI$_3$ with different stacking order, normalized to the saturation magnetization, as a function of an externally applied magnetic field. Arrows indicate the direction of the curve for increasing and decreasing fields.}
  \label{fig:hysteresis}
\end{figure}

\emph{No magnetic field} --- In the abscence of an external magnetic field, our LLG simulations show that a FM interlayer ordering is a (meta)stable state for each of the three stacking orders of the bilayer. However, an AFM ordering is only stable for the AB' and AA-stackings. In the AB-stacking, the structure always relaxes to a FM ordered state, even when the structure is initialized with an AFM one. This spontaneous change in magnetization is attributed to the large energy difference between FM and AFM phases. Hence, since it's dynamically unstable, the AFM-ordered AB-stacking will not be investigated any further.

\emph{Magnetic field effect on FM states} --- Applying an external magentic field to FM systems leads to hysteresis, as shown in Figure \ref{fig:hysteresis} for both monolayer and bilayer CrI$_3$. The hysteresis curve for the monolayer exhibits a very large coercive field of B = 8.5 T. For the FM-ordered AA, AB and AB' stackings, we find similar hysteresis curves also with coercive fields of respectively 7.5 T, 7.0 T, and 2.5 T. As discussed in the main text, flipping the magnetization by applying a magnetic field with opposite polarity will also change the sign of the Chern number, thus, reversing the propagation direction of the magnonic edge modes.

\begin{figure}[!ht]
  \centering
  \subfloat[][AA-stacking]{\includegraphics[width=.32\textwidth]{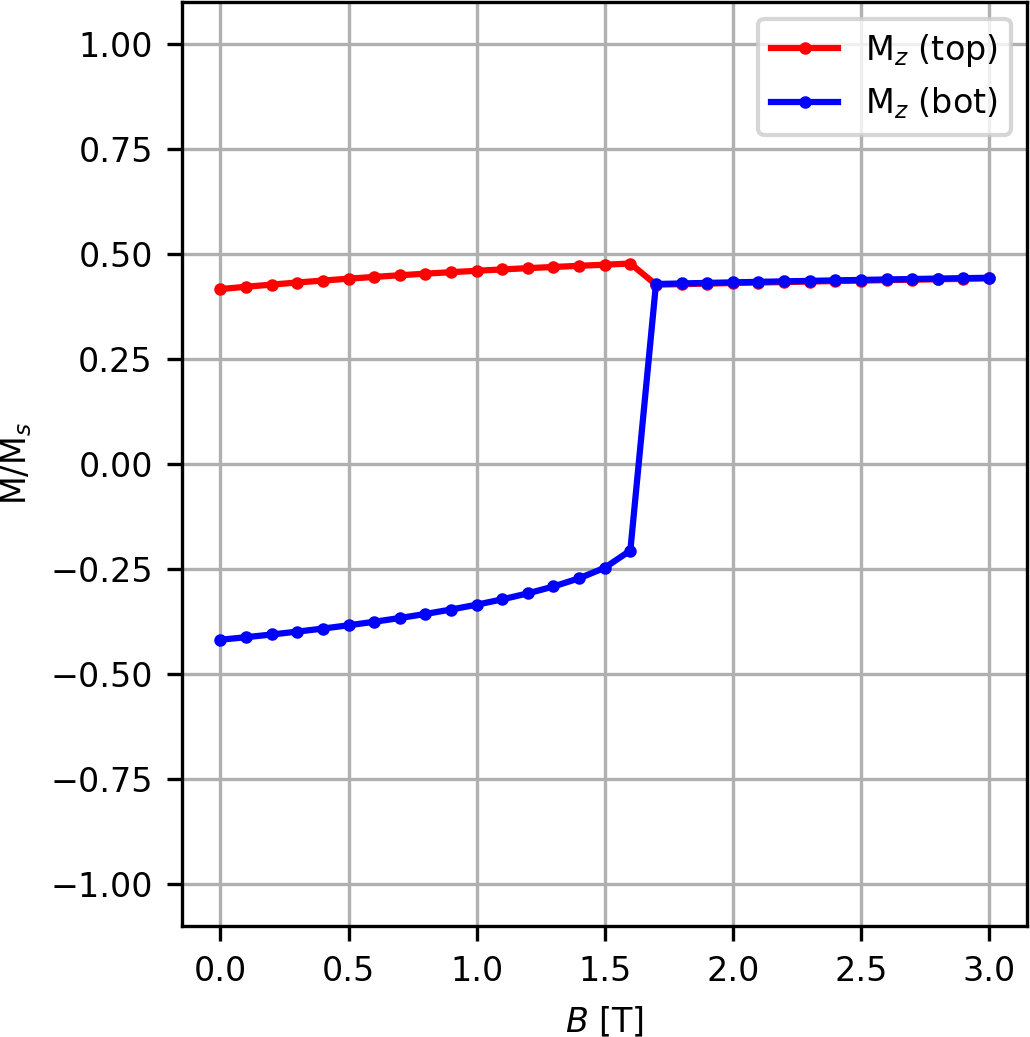}}\hfill
  \subfloat[][AB-stacking]{\includegraphics[width=.32\textwidth]{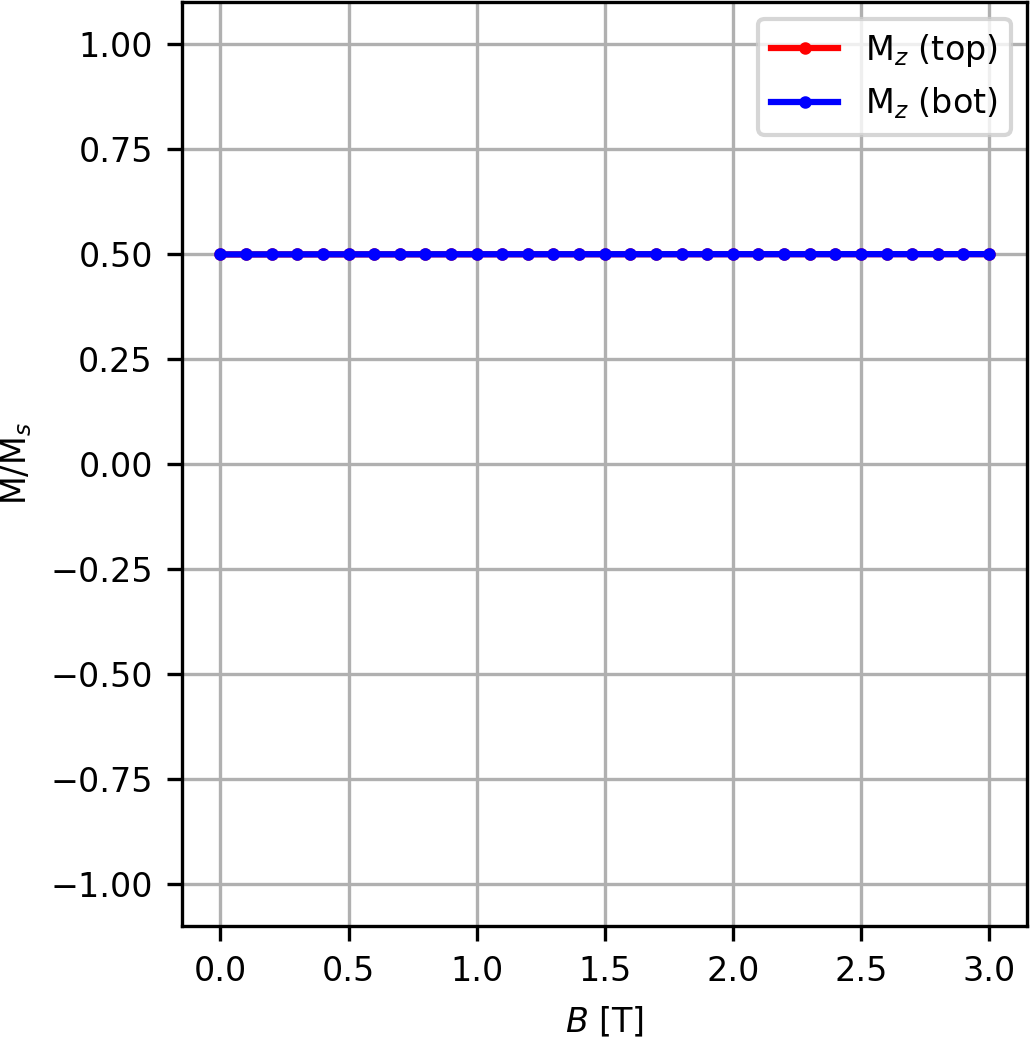}}\hfill
  \subfloat[][AB'-stacking]{\includegraphics[width=.32\textwidth]{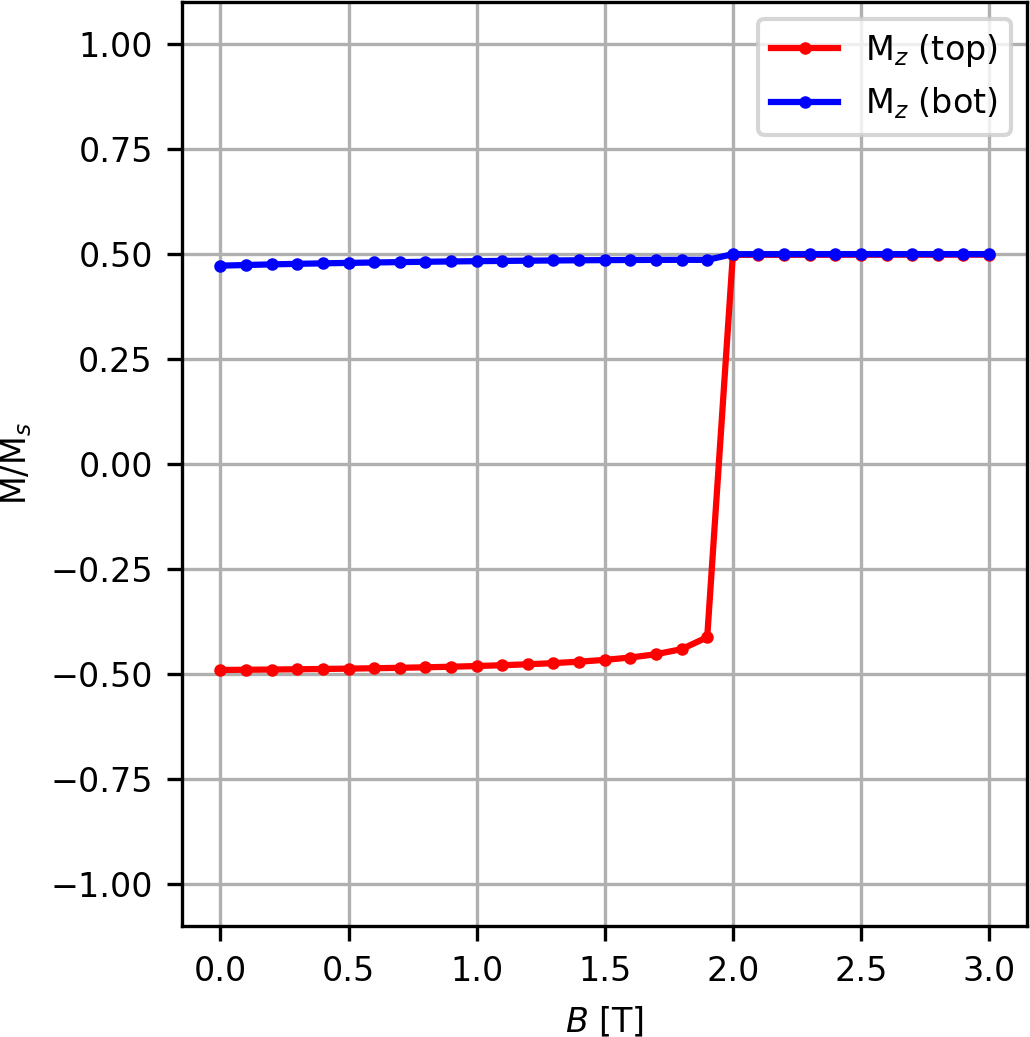}}
  \caption{\textbf{Magnetic field dependence of the interlayer magnetic ordering in the AFM ordered CrI$\mathbf{_3}$ bilayer for different stacking orders.} Magnetization of the top layer (red) and bottom layer (blue) of an AFM ordered CrI$_3$ bilayer as a function of the applied magnetic field, for respectively the AA-stacking (a), AB-stacking (b) and the AB'-stacking (c). The magnetization is normalized with respect to the total saturation magnetization.}
  \label{fig:sup_stacking_field}
\end{figure}

\emph{Magnetic field effect on AFM states} --- In Figure \ref{fig:sup_stacking_field}, we report results for the magnetic field dependence of the AFM interlayer ordering in bilayer CrI$_3$, by plotting the magnetization for the top and bottom layer separately as a function of the applied field. In the AB-stacking [Fig. \ref{fig:sup_stacking_field} (b)], the magnetization will always be fully saturated, even in the absence of a field, serving as evidence that an AFM interlayer ordering is probably not stable for this stacking. However, for the AA-stacking [Fig. \ref{fig:sup_stacking_field} (a)] and the AB'-stacking [Fig. \ref{fig:sup_stacking_field} (c)], we see that the AFM ordering remains stable up to relatively high fields of respectively B = 1.6 T and B = 1.9 T. For higher applied fields, the magnetization in one of the layers will flip (the layer with opposite magnetization to the field), leading to a FM ordering oriented parallel to the field. For a field with opposite orientation, we find similar results. As discussed in the main text, the applied magnetic field has some interesting effect on the magnonic properties, tuning both the size and topology of the bandgap in the AB'-stacking and shifting the position of the Dirac node upwards/downwards in the AA-stacking. 

\clearpage

\newpage
\section{Magnetic parameters}\label{supsec:parameters}
\begin{figure}[!b]
  \centering
  \includegraphics[width=.85\textwidth]{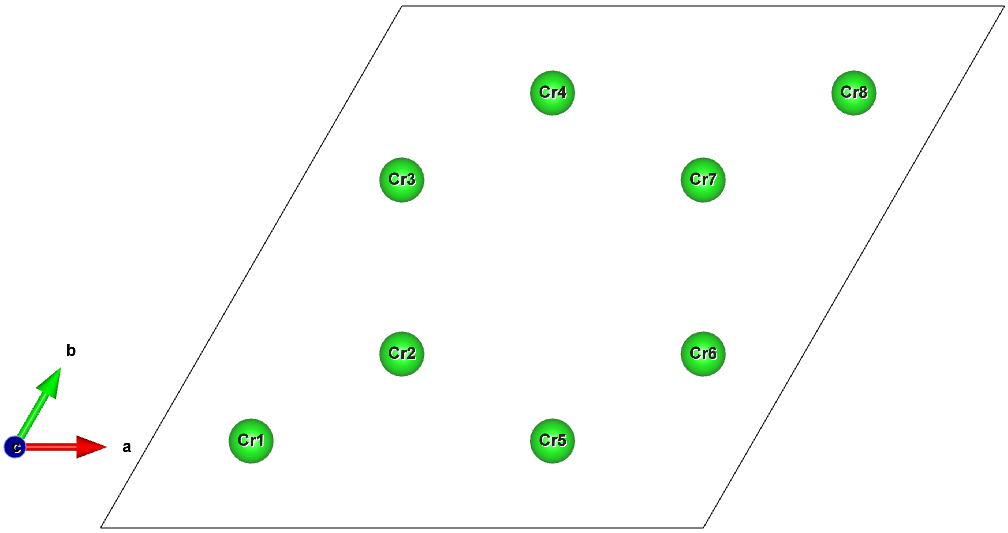}
  \caption{\textbf{Labeled spin sites in monolayer CrI$\mathbf{_3}$.} Top view of a $2 \times 2 \times 1$ supercell of monolayer CrI$_3$. Chromium atoms are represented by green spheres. Iodine atoms and bonds are not shown in the picture for the sake of simplicity. The supercell is marked with solid black lines. The crystal structures were drawn using VESTA\cite{VESTA}.}
  \label{fig:sup_mono}
\end{figure}

\begin{figure}[!b]
  \centering
  \subfloat[][Top view.]{\includegraphics[width=.5\textwidth]{figures_sup/aa1.png}}
  \subfloat[][Side view.]{\includegraphics[width=.5\textwidth]{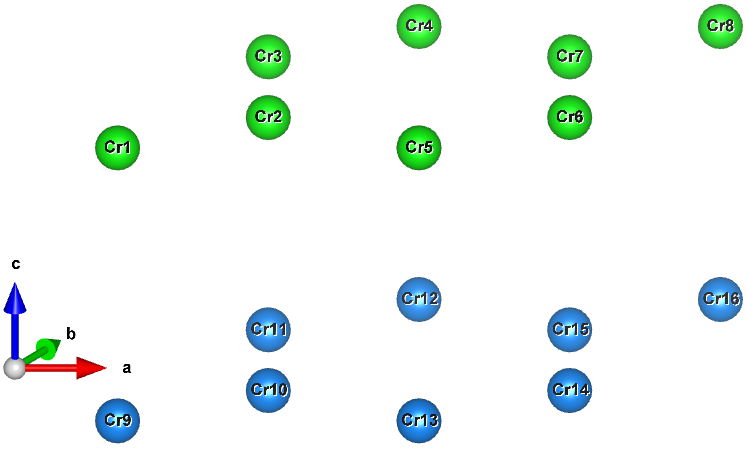}}
  \caption{\textbf{Labeled spin sites in an AA-stacked CrI$\mathbf{_3}$ bilayer.} Top view (a) and side view (b) of a $2 \times 2 \times 1$ supercell of an AA-stacked CrI$_3$ bilayer. Chromium atoms in the top- and bottom layer are represented by green and blue spheres respectively. Iodine atoms and bonds are not shown in the picture for the sake of simplicity. The supercell is marked with solid black lines in panel (a). The crystal structures were drawn using VESTA\cite{VESTA}.}
  \label{fig:sup_aa}
\end{figure}

\begin{figure}[!b]
  \centering
  \subfloat[][Top view.]{\includegraphics[width=.5\textwidth]{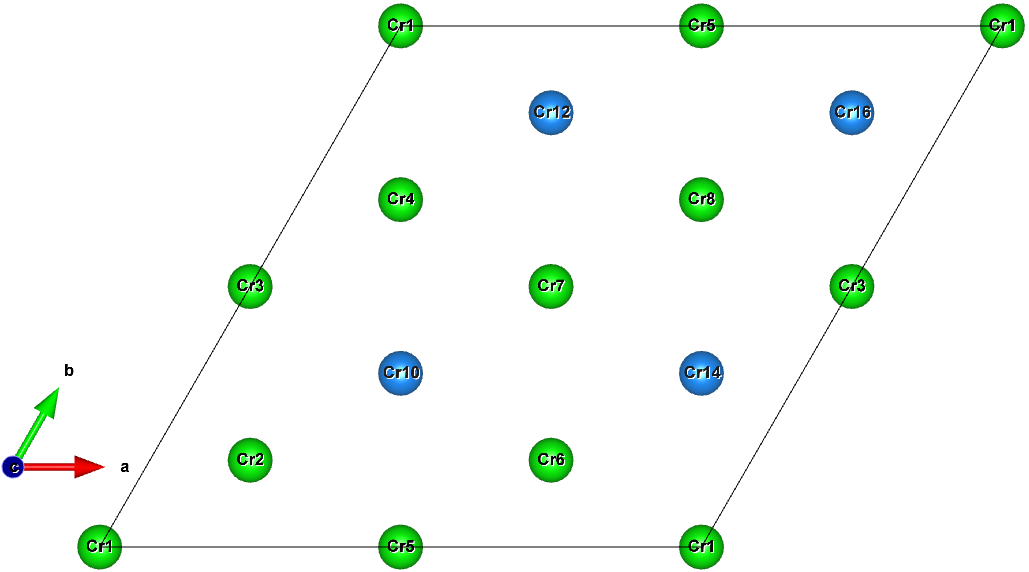}}
  \subfloat[][Side view.]{\includegraphics[width=.5\textwidth]{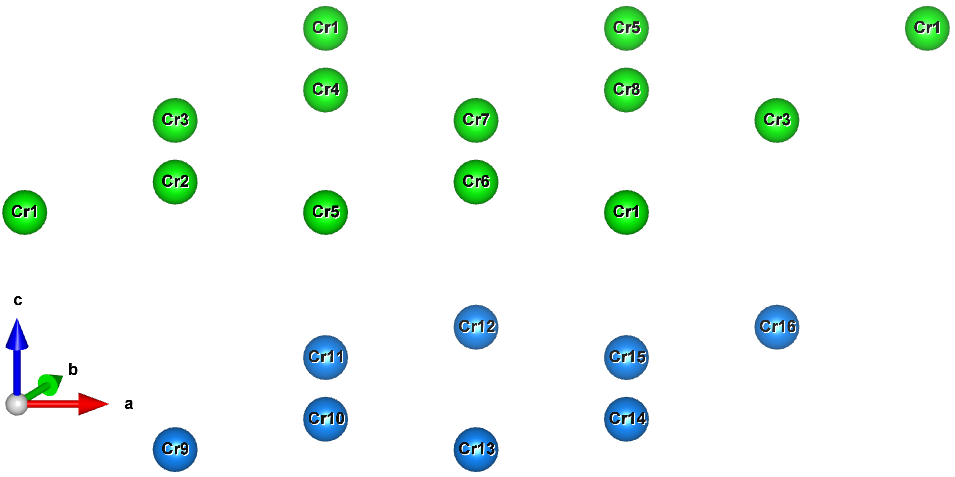}}
  \caption{\textbf{Labeled spin sites in an AB-stacked CrI$\mathbf{_3}$ bilayer.} Top view (a) and side view (b) of a $2 \times 2 \times 1$ supercell of an AB-stacked CrI$_3$ bilayer. Chromium atoms in the top- and bottom layer are represented by green and blue spheres respectively. Iodine atoms and bonds are not shown in the picture for the sake of simplicity. The supercell is marked with solid black lines in panel (a). The crystal structures were drawn using VESTA\cite{VESTA}.}
  \label{fig:sup_ab}
\end{figure}

\begin{figure}[!t]
  \centering
  \subfloat[][Top view.]{\includegraphics[width=0.95\textwidth]{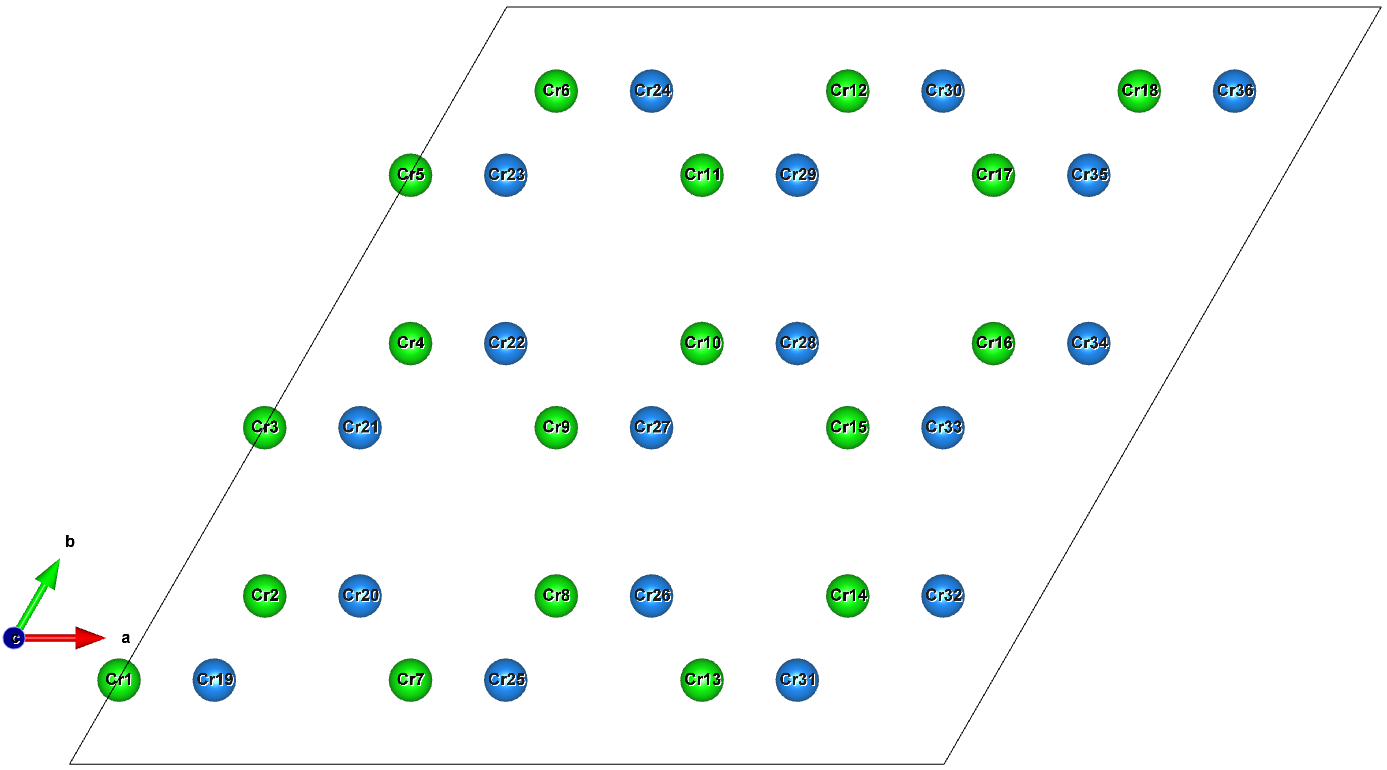}}\\
  \subfloat[][Side view.]{\includegraphics[width=0.85\textwidth]{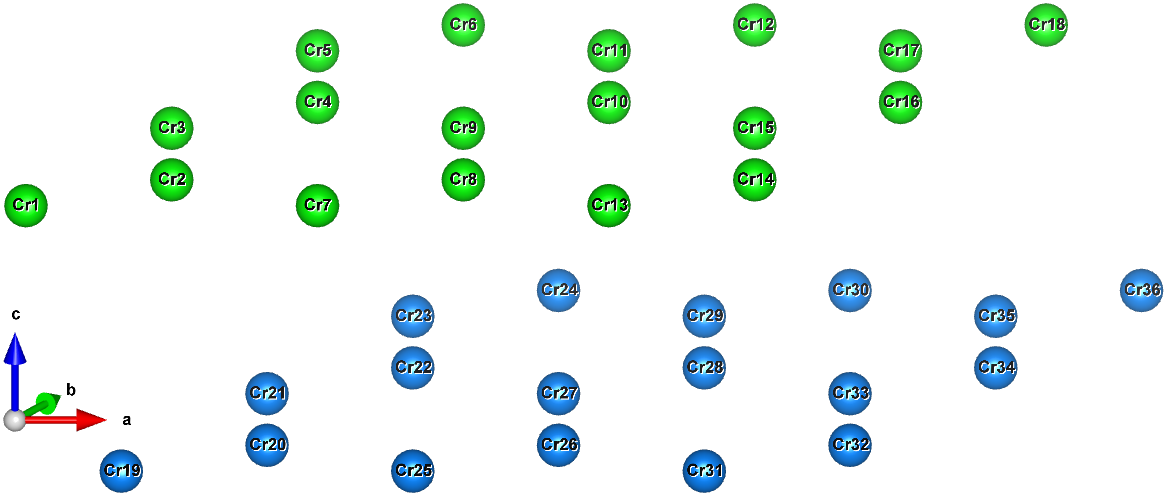}}
  \caption{\textbf{Labeled spin sites in an AB'-stacked CrI$\mathbf{_3}$ bilayer.} Top view (a) and side view (b) of a $3 \times 3 \times 1$ supercell of an AB'-stacked CrI$_3$ bilayer. Chromium atoms in the top- and bottom layer are represented by green and blue spheres respectively. Iodine atoms and bonds are not shown in the picture for the sake of simplicity. The supercell is marked with solid black lines in panel (a). The crystal structures were drawn using VESTA\cite{VESTA}.}
  \label{fig:sup_abb}
\end{figure}

In this section, we present the magnetic parameters -- obtained through a 4SM analysis -- for all the pairs necessary to characterize the magnonic behavior of monolayer and bilayer CrI$_3$. Tables \ref{tab:sup_parameters_mono} - \ref{tab:sup_parameters_abb}, contain the magnetic parameters for respectively the monolayer [Fig. \ref{fig:sup_mono}], the AA-stacking [Fig. \ref{fig:sup_aa}], the AB-stacking [Fig. \ref{fig:sup_ab}], and the AB'-stacking [Fig. \ref{fig:sup_abb}]. For these four systems, we report parameters for the NN and NNN intra- and interlayer exchange and DMI [Tab. \ref{tab:sup_parameters_mono} - \ref{tab:sup_parameters_abb}], the SIA, and the Kitaev interactions [Tab. \ref{tab:sup_para_kitaev}]. The exchange- and DMI parameters are reported in Cartesian coordinates [Tab. \ref{tab:sup_parameters_mono} - \ref{tab:sup_parameters_abb}], afterwards in Table \ref{tab:sup_para_kitaev}, we consider local eigenbases that diagonalize the symmetric exchange matrices to quantify the Kitaev interactions. 

In all considered structures, the intralayer exchange is FM and anisotropic with the NN interaction delivering the dominant contribution, the NNN exchange also prefers a FM ordering but is weaker than the former. The out-of-plane exchange anisotropy, i.e. $\Delta = \langle J_{ij}^{zz}\rangle - \langle J_{ij}^{\alpha \alpha}\rangle$ with $\alpha = \{x,y\}$, is the most important contribution to the magnetic anisotropy of CrI$_3$ and causes the spins to prefer an out-of-plane orientation. In monolayer CrI$_3$, the NN DMI is equal to zero due to the material's inversion symmetry. Stacking two layers in a bilayer breaks this inversion symmetry leading to a non-zero NN DMI in all stacking orders. Between intralayer NNN spins, there is no inversion symmetry resulting, for all structures, in a small non-zero DMI.

The AB- and AA-stacked bilayers, both show a FM NN and NNN interlayer coupling. The strength of the interlayer coupling is significantly stronger for the AB-stacking as it has more interacting pairs that also interact more strongly. In the AB'-stacked bilayer, there is competition between the FM NN interlayer coupling and the AFM NNN and 3NN interlayer coupling, resulting in an overall preference for an AFM interlayer ordering. For all stackings, the interlayer DMI is significantly weaker than the intralayer one, having almost no influence on the material's properties. 

Notice that, in this work, we only report 3NN parameters for the AB'-stacking. In this stacking, the 3NN interlayer interaction is indispensable in order to achieve an overall AFM interlayer ordering as the ground state, in agreement with experiment\cite{huang2017,thiel2019,song2019,li2019,chen2019}. In all other cases, we can safely neglect the 3NN interactions since they deliver only minor contributions to the overall exchange. To illustrate the latter, we calculate several 3NN parameters to demonstrate that their characteristic orders of magnitude are negligible. For example, in the AB-stacked bilayer we found 3NN intralayer exchange pairs of $(J^{xx}_{2-7},J^{yy}_{2-7},J^{zz}_{2-7}) = (0.00, 0.00, 0.02)$ meV. Similarly, we find a value of $J^{zz}_{2-7}$ = 0.03 meV for the monolayer\footnote[1]{Note that these 3NN parameters have been calculated for supercells with increasing sizes (2$\times$2$\times$1, 3$\times$3$\times$1, 4$\times$4$\times$1), to verify that its value is converged as a function of the supercell size.}. Hence, we can safely assume that the 3NN intralayer interaction is negligible in all structures. Regarding the 3NN interlayer exchange, we found values of $(J^{xx}_{7-16},J^{yy}_{7-16},J^{zz}_{7-16}) = (0.01, 0.00, 0.00)$ meV in the AB- and AA-stacking, values like this can also safely be neglected without significant influence on further results. 

In principle, the SIA matrix is symmetric resulting in six unique matrix elements that need to be calculated. However, notice that, when there is a 3-, 4-, or 6-fold rotational symmetry around the out-of-plane axis, it suffices to calculate only one parameter $A_{ii}^{zz}$ \cite{sabani2020}. For the monolayer, the AA-stacking, and the AB-stacking, in which this symmetry is upheld, we find values for $\langle A_{ii}^{zz}\rangle$ of respectively -0.08 meV, -0.08 meV and -0.07 meV. In the AB'-stacked bilayer, where this symmetry is absent, we need to calculate the full SIA-matrix, however, all elements vanish except $\langle A_{ii}^{zz}\rangle$ = -0.07 meV and $\langle A^{yz}_{ii}\rangle$ = $\langle A^{zy}_{ii}\rangle$ = 0.02 meV. Hence, for all structures, the SIA shows a preference for an out-of-plane orientation of the spins, delivering a small contribution to the material's magnetic anisotropy.

To calculate the Kitaev constants, we consider for each spin pair, an eigenbasis $\{\lambda,\mu,\nu\}$ that diagonalizes the corresponding symmetric exchange matrix, yielding three eigenvalues $J_{ij}^\lambda$, $J_{ij}^\mu$, and $J_{ij}^\nu$. In the case that $J_{ij}^\lambda \approx J_{ij}^\mu \neq J_{ij}^\nu$, the exchange constant is defined as $\mathcal{J}_{ij} = (J_{ij}^\lambda + J_{ij}^\mu)/2$ and the corresponding Kitaev constant as $\mathcal{K}_{ij} = J_{ij}^\nu - \mathcal{J}_{ij}$. The DMI in the local basis is found by projecting the Cartesian DMI vector along the $\{\lambda,\mu,\nu\}$ directions. We summarize the exchange and Kitaev constants in Table \ref{tab:sup_para_kitaev}. In all structures, the main contribution to the Kitaev interaction comes from the NN intralayer exchange. 

\newpage
\begin{table}[!t]
\caption{\label{tab:sup_parameters_mono}\textbf{Exchange parameters for monolayer CrI$_\mathbf{3}$.} For each interacting pair of spins, the table contains the elements of both the NN and the NNN intralayer exchange matrices $J_{ij}$, the out-of-plane exchange anisotropy $\Delta$, and the components of the corresponding DMI-vectors $\mathbf{D}_{ij}$. All parameters are given in Cartesian coordinates. Corresponding pairs are indicated in Figure \ref{fig:sup_mono}.}
\begin{tabular}{lc||ccc|c|cccccc||ccc|c}
\hline\hline
   & pair & $J^{xx}_{ij}$ & $J^{yy}_{ij}$ & $J^{zz}_{ij}$ & $\Delta$ & $J^{xy}_{ij}$ & $ J^{yx}_{ij}$ & $J^{xz}_{ij}$ & $J^{zx}_{ij}$ & $ J^{yz}_{ij}$ & $J^{zy}_{ij}$ & $|D^{x}_{ij}|$ & $|D^{y}_{ij}|$ & $|D^{z}_{ij}|$ & $|\mathbf{D}_{ij}|$ \\
   & (i-j) & (meV) & (meV) & (meV) & (meV) & (meV) & (meV) & (meV) & (meV) & (meV) & (meV) & (meV) & (meV) & (meV) & (meV) \\ \hline
 NN & (1-2) & -4.34 & -3.24 & -3.96 & & 0.00 & 0.00 & 0.00 & 0.00 & -0.65 & -0.65 & 0.00 & 0.00 & 0.00 & 0.00 \\
          & (2-3) & -3.52 & -4.07 & -3.96 & & -0.48 & -0.48 & -0.56 & -0.56 & 0.33 & 0.33 & 0.00 & 0.00 & 0.00 & 0.00 \\
          & (2-5) & -3.52 & -4.07 & -3.96 & & 0.48 & 0.48 & 0.56 & 0.56 & 0.33 & 0.33 & 0.00 & 0.00 & 0.00 & 0.00 \\  
          & $\langle J_{ij}\rangle$ & -3.79 & -3.79 & -3.96 & -0.17 & 0.00 & 0.00 & 0.00 & 0.00 & 0.00 & 0.00 & 0.00 & 0.00 & 0.00 & 0.00 \\
          & & & & & & & & & & & & & & \\
       NNN & (1-3) & -0.67 & -0.73 & -0.66 & & 0.08 & 0.03 & 0.09 & 0.03 & -0.01 & 0.09 & 0.05 & 0.03 & 0.03 & 0.06 \\
           & (1-5) & -0.67 & -0.73 & -0.66 & & -0.08 & -0.03 & -0.09 & -0.03 & -0.01 & 0.09 & 0.05 & 0.03 & 0.03 & 0.06 \\
           & (3-5) & -0.76 & -0.64 & -0.66 & & 0.03 & -0.03 & -0.05 & 0.05 & -0.08 & -0.08 & 0.00 & 0.05 & 0.03 & 0.06 \\
          & $\langle J_{ij}\rangle$ & -0.70 & -0.70 & -0.66 & 0.04 & 0.01 & -0.01 & -0.02 & 0.02 & -0.03 & 0.03 & 0.03 & 0.04 & 0.03 & 0.06 \\ \hline \hline
\end{tabular}
\end{table}

\clearpage

\newpage
\begin{table}[!t]
\caption{\label{tab:sup_parameters_aa} \textbf{Exchange parameters for an AA-stacked CrI$_\mathbf{3}$ bilayer.} For each interacting pair of spins, the table contains the elements of both the NN and the NNN intra- and interlayer exchange matrices $J_{ij}$, the out-of-plane exchange anisotropy $\Delta$, and the components of the corresponding DMI-vectors $\mathbf{D}_{ij}$. All parameters are given in Cartesian coordinates. Corresponding pairs are indicated in Figure \ref{fig:sup_aa}.}
\begin{tabular}{lc||ccc|c|cccccc||ccc|c}
\hline \hline
   & pair & $J^{xx}_{ij}$ & $J^{yy}_{ij}$ & $J^{zz}_{ij}$ & $\Delta$ & $J^{xy}_{ij}$ & $ J^{yx}_{ij}$ & $J^{xz}_{ij}$ & $J^{zx}_{ij}$ & $ J^{yz}_{ij}$ & $J^{zy}_{ij}$ & $|D^{x}_{ij}|$ & $|D^{y}_{ij}|$ & $|D^{z}_{ij}|$ & $|\mathbf{D}_{ij}|$ \\
   & (i-j) & (meV) & (meV) & (meV) & (meV) & (meV) & (meV) & (meV) & (meV) & (meV) & (meV) & (meV) & (meV) & (meV) & (meV) \\ \hline
   \textbf{Intra} & & & & & & & & & & & & & & \\
 NN & (1-2) & -4.40 & -3.34 & -4.05 & & 0.04 & -0.04 & 0.11 & -0.11 & -0.63 & -0.63 & 0.00 & 0.11 & 0.04 & 0.12 \\
    & (2-3) & -3.62 & -4.15 & -4.06 & & -0.52 & -0.44 & -0.49 & -0.60 & 0.41 & 0.22 & 0.10 & 0.06 & 0.04 & 0.12 \\
    & (2-5) & -3.62 & -4.15 & -4.06 & & 0.44 & 0.52 & 0.60 & 0.49 & 0.22 & 0.41 & 0.10 & 0.05 & 0.04 & 0.12 \\
    & $\langle J_{ij}\rangle$ & -3.88 & -3.88 & -4.05 & -0.17 & -0.01 & 0.01 & 0.07 & -0.07 & 0.00 & 0.00 & 0.07 & 0.07 & 0.04 & 0.12 \\
    & & & & & & & & & & & & & & \\
    NNN & (1-3) & -0.59 & -0.64 & -0.57 & & 0.10 & 0.02 & 0.08 & 0.03 & 0.00 & 0.06 & 0.03 & 0.02 & 0.04 & 0.06 \\
    & (1-5) & -0.67 & -0.56 & -0.57 & & 0.04 & -0.05 & -0.04 & 0.04 & -0.07 & -0.06 & 0.00 & 0.04 & 0.04 & 0.06 \\
    & (3-5) & -0.59 & -0.65 & -0.57 & & -0.09 & -0.01 & -0.08 & -0.03 & 0.00 & 0.07 & 0.04 & 0.02 & 0.04 & 0.06 \\
    & $\langle J_{ij}\rangle$ & -0.62 & -0.62 & -0.57 & 0.05 & 0.02 & -0.01 & -0.01 & 0.01 & -0.02 & 0.02 & 0.03 & 0.03 & 0.04 & 0.06 \\ \hline
       \textbf{Inter} & & & & & & & & & & & & & & \\
NN & (1-9) & -0.12 & -0.12 & -0.04 & 0.08 & -0.06 & 0.06 & 0.00 & 0.00 & 0.00 & 0.00 & 0.00 & 0.00 & 0.06 & 0.06 \\
& & & & & & & & & & & & & & \\
           NNN & (7-12) & -0.13 & -0.11 & -0.12 & & -0.01 & -0.01 & 0.00 & 0.00 & -0.01 & -0.01 & 0.00 & 0.00 & 0.00 & 0.00 \\
           & (7-14) & -0.13 & -0.12 & -0.12 & & 0.01 & 0.01 & 0.01 & 0.01 & 0.01 & 0.01 & 0.00 & 0.00 & 0.00 & 0.00 \\ 
           & (7-16) & -0.11 & -0.13 & -0.11 & & -0.01 & -0.01 & -0.01 & -0.01 & 0.00 & 0.00 & 0.00 & 0.00 & 0.00 & 0.00 \\ 
    & $\langle J_{ij}\rangle$ & -0.12 & -0.12 & -0.12 & 0.00 & 0.00 & 0.00 & 0.00 & 0.00 & 0.00 & 0.00 & 0.00 & 0.00 & 0.00 & 0.00 \\ \hline \hline
\end{tabular}
\end{table}

\clearpage

\newpage
\begin{table}[!t]
\caption{\label{tab:sup_parameters_ab} \textbf{Exchange parameters for an AB-stacked CrI$_\mathbf{3}$ bilayer.} For each interacting pair of spins, the table contains the elements of both the NN and the NNN intra- and interlayer exchange matrices $J_{ij}$, the out-of-plane exchange anisotropy $\Delta$, and the components of the corresponding DMI-vectors $\mathbf{D}_{ij}$. Since the sublattice symmetry is broken, parameters are given for the two sublattices (A,B) separately. All parameters are given in Cartesian coordinates. Corresponding pairs are indicated in Figure \ref{fig:sup_ab}.}
\begin{tabular}{lc||ccc|c|cccccc||ccc|c}
\hline \hline
   & pair & $J^{xx}_{ij}$ & $J^{yy}_{ij}$ & $J^{zz}_{ij}$ & $\Delta$ & $J^{xy}_{ij}$ & $ J^{yx}_{ij}$ & $J^{xz}_{ij}$ & $J^{zx}_{ij}$ & $ J^{yz}_{ij}$ & $J^{zy}_{ij}$ & $|D^{x}_{ij}|$ & $|D^{y}_{ij}|$ & $|D^{z}_{ij}|$ & $|\mathbf{D}_{ij}|$ \\
   & (i-j) & (meV) & (meV) & (meV) & (meV) & (meV) & (meV) & (meV) & (meV) & (meV) & (meV) & (meV) & (meV) & (meV) & (meV) \\ \hline
   \textbf{Intra} & & & & & & & & & & & & & & \\
NN  & (1-2) & -4.63 & -3.56 & -4.25 & & -0.04 & 0.03 & -0.13 & 0.11 & -0.64 & -0.64 & 0.00 & 0.12 & 0.03 & 0.13 \\
    & (2-3) & -3.43 & -3.93 & -3.86 & & -0.47 & -0.43 & -0.54 & -0.50 & 0.26 & 0.34 & 0.04 & 0.02 & 0.02 & 0.05 \\
    & (2-5) & -3.84 & -4.34 & -4.25 & & 0.49 & 0.53 & 0.50 & 0.61 & 0.42 & 0.22 & 0.10 & 0.05 & 0.02 & 0.11 \\
    & $\langle J_{ij}\rangle$ & -3.96 & -3.94 & -4.12 & -0.16 & -0.01 & 0.04 & -0.06 & 0.07 & 0.02 & -0.02 & 0.05 & 0.06 & 0.02 & 0.10 \\
    NNN$_\mathrm{A}$ & (1-3) & -0.53 & -0.58 & -0.52 & & 0.08 & 0.02 & 0.07 & 0.02 & 0.02 & 0.06 & 0.02 & 0.02 & 0.03 & 0.04 \\
    & (1-5) & -0.53 & -0.58 & -0.51 & & -0.08 & -0.02 & -0.07 & -0.03 & -0.01 & 0.05 & 0.03 & 0.02 & 0.03 & 0.05 \\
    & (3-5) & -0.61 & -0.50 & -0.50 & & 0.03 & -0.03 & -0.04 & 0.04 & -0.06 & -0.06 & 0.00 & 0.04 & 0.03 & 0.05 \\
    & $\langle J_{ij}\rangle$ & -0.55 & -0.55 & -0.51 & 0.04 & 0.01 & -0.01 & -0.01 & 0.01 & -0.02 & 0.02 & 0.02 & 0.03 & 0.03 & 0.05 \\
    NNN$_\mathrm{B}$ & (2-4) & -0.58 & -0.64 & -0.57 & & 0.02 & 0.08 & 0.03 & 0.08 & 0.07 & 0.00 & 0.04 & 0.03 & 0.03 & 0.06 \\
    & (2-6) & -0.67 & -0.56 & -0.57 & & -0.03 & 0.04 & 0.04 & -0.04 & -0.06 & -0.07 & 0.00 & 0.04 & 0.03 & 0.05 \\
    & (4-6) & -0.59 & -0.64 & -0.57 & & -0.02 & -0.09 & -0.03 & -0.08 & 0.07 & -0.01 & 0.04 & 0.02 & 0.03 & 0.06 \\
    & $\langle J_{ij}\rangle$ & -0.61 & -0.61 & -0.57 & 0.04 & -0.01 & 0.01 & 0.01 & -0.01 & 0.03 & -0.02 & 0.03 & 0.03 & 0.03 & 0.05 \\ \hline
       \textbf{Inter} & & & & & & & & & & & & & & \\
NN$_\mathrm{B}$ & (2-9) & -0.26 & -0.27 & -0.30 & -0.02 & 0.06 & 0.06 & 0.01 & 0.02 & 0.01 & 0.01 & 0.00 & 0.00 & 0.00 & 0.00 \\
           NNN$_\mathrm{A}$ & (7-10) & -0.39 & -0.49 & -0.46 & & -0.04 & -0.04 & -0.04 & -0.04 & 0.05 & 0.05 & 0.00 & 0.00 & 0.00 & 0.00 \\ 
           & (7-11) & -0.31 & -0.30 & -0.32 & & -0.03 & -0.03 & -0.02 & -0.01 & -0.01 & -0.01 & 0.00 & 0.00 & 0.00 & 0.01 \\ 
           & (7-12) & -0.49 & -0.39 & -0.47 & & -0.02 & -0.02 & -0.02 & -0.02 & -0.06 & -0.06 & 0.00 & 0.00 & 0.00 & 0.00 \\ 
           & (7-13) & -0.32 & -0.28 & -0.31 & & 0.02 & 0.02 & 0.02 & 0.02 & -0.02 & 0.00 & 0.01 & 0.00 & 0.00 & 0.01 \\ 
           & (7-14) & -0.43 & -0.45 & -0.47 & & 0.06 & 0.06 & 0.06 & 0.06 & 0.01 & 0.01 & 0.00 & 0.00 & 0.00 & 0.00 \\ 
           & (7-15) & -0.28 & -0.33 & -0.31 & & 0.01 & 0.01 & 0.01 & 0.00 & 0.03 & 0.02 & 0.00 & 0.01 & 0.00 & 0.01 \\
    & $\langle J_{ij}\rangle$ & -0.37 & -0.37 & -0.39 & -0.02 & 0.00 & 0.00 & 0.00 & 0.00 & 0.00 & 0.00 & 0.00 & 0.00 & 0.00 & 0.01 \\
           NNN$_\mathrm{B}$ & (8-12) & -0.31 & -0.30 & -0.32 & & -0.03 & -0.03 & -0.01 & -0.02 & -0.01 & -0.01 & 0.00 & 0.01 & 0.00 & 0.01 \\
           & (8-14) & -0.32 & -0.28 & -0.31 & & 0.02 & 0.02 & 0.02 & 0.02 & 0.00 & -0.02 & 0.01 & 0.00 & 0.00 & 0.01 \\ 
           & (8-16) & -0.28 & -0.33 & -0.31 & & 0.01 & 0.01 & 0.00 & 0.01 & 0.02 & 0.03 & 0.00 & 0.01 & 0.00 & 0.01 \\ 
    & $\langle J_{ij}\rangle$ & -0.30 & -0.30 & -0.32 & -0.02 & 0.00 & 0.00 & 0.00 & 0.00 & 0.00 & 0.00 & 0.00 & 0.01 & 0.00 & 0.01 \\ \hline \hline
\end{tabular}
\end{table}

\clearpage

\newpage
\begin{table}[!t]
\caption{\label{tab:sup_parameters_abb} \textbf{Exchange parameters for an AB'-stacked CrI$_\mathbf{3}$ bilayer.} For each interacting pair of spins, the table contains the elements of both the NN and the NNN intra- and interlayer exchange matrices $J_{ij}$, the out-of-plane exchange anisotropy $\Delta$, and the components of the corresponding DMI-vectors $\mathbf{D}_{ij}$. Since the sublattice symmetry is broken, parameters are given for the two sublattices (A,B) separately. All parameters are given in Cartesian coordinates. Corresponding pairs are indicated in Figure \ref{fig:sup_abb}.}
\begin{tabular}{lc||ccc|c|cccccc||ccc|c}
\hline \hline
   & pair & $J^{xx}_{ij}$ & $J^{yy}_{ij}$ & $J^{zz}_{ij}$ & $\Delta$ & $J^{xy}_{ij}$ & $ J^{yx}_{ij}$ & $J^{xz}_{ij}$ & $J^{zx}_{ij}$ & $ J^{yz}_{ij}$ & $J^{zy}_{ij}$ & $|D^{x}_{ij}|$ & $|D^{y}_{ij}|$ & $|D^{z}_{ij}|$ & $|\mathbf{D}_{ij}|$ \\
   & (i-j) & (meV) & (meV) & (meV) & (meV) & (meV) & (meV) & (meV) & (meV) & (meV) & (meV) & (meV) & (meV) & (meV) & (meV) \\ \hline
   \textbf{Intra} & & & & & & & & & & & & & & \\
NN  & (1-2) & -4.63 & -3.54 & -4.25 & & 0.06 & -0.01 & -0.12 & 0.12 & -0.66 & -0.64 & 0.01 & 0.12 & 0.03 & 0.12 \\
    & (2-3) & -3.43 & -3.93 & -3.86 & & -0.48 & -0.44 & -0.54 & -0.50 & 0.26 & 0.34 & 0.04 & 0.02 & 0.02 & 0.05 \\
    & (2-7) & -3.85 & -4.34 & -4.26 & & 0.47 & 0.54 & 0.51 & 0.61 & 0.43 & 0.22 & 0.11 & 0.05 & 0.03 & 0.12 \\
    & $\langle J_{ij}\rangle$ & -3.97 & -3.94 & -4.12 & -0.15 & 0.02 & 0.03 & -0.05 & 0.08 & 0.01 & -0.03 & 0.05 & 0.06 & 0.03 & 0.10 \\
    & & & & & & & & & & & & & & \\
    NNN$_{\mathrm{A}}$ & (1-3) & -0.61 & -0.68 & -0.60 & & 0.08 & 0.02 & 0.08 & 0.04 & 0.00 & 0.07 & 0.03 & 0.02 & 0.03 & 0.05 \\
    & (1-7) & -0.64 & -0.53 & -0.54 & & 0.02 & -0.01 & -0.04 & 0.03 & -0.07 & -0.07 & 0.00 & 0.03 & 0.02 & 0.04 \\
    & (3-7) & -0.57 & -0.62 & -0.54 & & -0.10 & 0.00 & -0.06 & -0.04 & 0.02 & 0.05 & 0.02 & 0.01 & 0.05 & 0.05 \\
    & $\langle J_{ij}\rangle$ & -0.61 & -0.61 & -0.56 & 0.05 & 0.00 & 0.00 & -0.01 & 0.01 & -0.02 & 0.02 & 0.02 & 0.02 & 0.03 & 0.05 \\
    & & & & & & & & & & & & & & \\
    NNN$_{\mathrm{B}}$ & (2-4) & -0.56 & -0.61 & -0.54 & & 0.04 & 0.07 & 0.03 & 0.07 & 0.07 & 0.00 & 0.03 & 0.02 & 0.02 & 0.04 \\
    & (2-8) & -0.56 & -0.62 & -0.54 & & 0.00 & -0.10 & -0.05 & -0.07 & 0.04 & 0.01 & 0.02 & 0.01 & 0.05 & 0.05 \\
    & (4-8) & -0.70 & -0.58 & -0.59 & & -0.04 & 0.03 & 0.04 & -0.03 & -0.07 & -0.08 & 0.01 & 0.04 & 0.04 & 0.05 \\
    & $\langle J_{ij}\rangle$ & -0.61 & -0.60 & -0.56 & 0.05 & 0.00 & 0.00 & 0.01 & -0.01 & 0.01 & -0.02 & 0.02 & 0.02 & 0.03 & 0.05 \\ \hline
       \textbf{Inter} & & & & & & & & & & & & & & \\
NN & (7-20) & -0.30 & -0.20 & -0.30 &  & 0.03 & 0.03 & -0.01 & 0.00 & -0.02 & -0.02 & 0.00 & 0.00 & 0.00 & 0.00 \\
& (1-19) & -0.26 & -0.31 & -0.29 & & -0.03 & -0.06 & -0.05 & 0.00 & -0.02 & 0.06 & 0.04 & 0.03 & 0.01 & 0.05 \\
& (2-19) & -0.26 & -0.27 & -0.28 & & 0.05 & 0.05 & 0.02 & 0.02 & 0.02 & 0.02 & 0.00 & 0.00 & 0.00 & 0.00 \\
& (2-20) & -0.26 & -0.29 & -0.29 & & -0.05 & -0.04 & 0.01 & -0.05 & 0.06 & -0.03 & 0.04 & 0.03 & 0.01 & 0.05 \\
& & & & & & & & & & & & & & \\
    NNN & (9-21) & 0.15 & 0.12 & 0.13 & & -0.03 & -0.02 & 0.00 & -0.04 & 0.04 & -0.02 & 0.03 & 0.02 & 0.01 & 0.04 \\
        & (9-26) & 0.15 & 0.20 & 0.18 & & -0.01 & -0.01 & 0.00 & 0.00 & -0.02 & -0.02 & 0.00 & 0.00 & 0.00 & 0.00 \\
        & (10-22) & 0.15 & 0.12 & 0.13 & & -0.02 & -0.03 & -0.04 & 0.00 & -0.02 & 0.04 & 0.03 & 0.02 & 0.01 & 0.04 \\
        & (10-29) & 0.21 & 0.17 & 0.19 & & 0.02 & 0.02 & 0.02 & 0.02 & 0.00 & 0.00 & 0.00 & 0.00 & 0.00 & 0.00 \\ 
        & & & & & & & & & & & & & & \\
        3NN & (9-20) & 0.12 & 0.12 & 0.12 & & 0.00 & 0.00 & 0.00 & 0.00 & 0.00 & 0.00 & 0.00 & 0.00 & 0.00 & 0.00 \\
& (9-23) & 0.07 & 0.07 & 0.06 & & -0.03 & -0.02 & 0.00 & -0.04 & 0.04 & -0.02 & 0.03 & 0.02 & 0.01 & 0.04 \\ 
& (9-25) & 0.06 & 0.06 & 0.06 & & -0.06 & 0.03 & -0.08 & 0.08 & -0.01 & 0.00 & 0.01 & 0.08 & 0.05 & 0.09 \\
& (9-28) & 0.02 & 0.01 & 0.01 & & -0.01 & -0.01 & 0.00 & 0.00 & 0.00 & 0.00 & 0.00 & 0.00 & 0.00 & 0.00 \\ 
& $\langle J_{ij}\rangle$ & 0.07 & 0.07 & 0.07 & & -0.03 & 0.00 & -0.02 & 0.01 & 0.01 & 0.00 & 0.01 & 0.02 & 0.01 & 0.03 \\
& & & & & & & & & & & & & & \\
& (10-23) & 0.13 & 0.13 & 0.12 & & 0.00 & 0.00 & 0.00 & 0.00 & 0.00 & 0.00 & 0.00 & 0.00 & 0.00 & 0.00 \\
& (10-24) & 0.07 & 0.07 & 0.06 & & 0.03 & -0.03 & -0.03 & 0.03 & 0.08 & -0.07 & 0.07 & 0.03 & 0.03 & 0.09 \\
& (10-26) & 0.06 & 0.07 & 0.05 & & 0.03 & -0.03 & 0.08 & -0.08 & 0.01 & -0.01 & 0.01 & 0.08 & 0.03 & 0.09 \\
& (10-33) & 0.02 & 0.01 & 0.01 & & -0.01 & -0.01 & 0.00 & 0.00 & 0.00 & 0.00 & 0.00 & 0.00 & 0.00 & 0.00 \\
& $\langle J_{ij}\rangle$ & 0.07 & 0.07 & 0.06 & & 0.01 & -0.01 & 0.01 & -0.01 & 0.02 & -0.02 & 0.02 & 0.03 & 0.02 & 0.03 \\ \hline \hline
\end{tabular}
\end{table}

\clearpage

\newpage
\begin{table}[!ht]
\caption{\label{tab:sup_para_kitaev}\textbf{Exchange parameters for CrI$_\mathbf{3}$ in the $\{\lambda,\mu,\nu\}$ basis.} Intra- and interlayer NN and NNN exchange, DMI and Kitaev parameters for monolayer (1L) and bilayer (2L) CrI$_3$. The Kitaev interaction for the AB-stacking and the AB'-stacking is anisotropic so multiple values are given for each interaction.}
\begin{tabular}{l @{\hskip 0.5cm} l @{\hskip 0.5cm} c @{\hskip 0.5cm}||c @{\hskip 1cm} c @{\hskip 1cm} c}
\hline\hline
   & & pair(s) & $\mathcal{J}_{ij}$ & $\mathcal{K}_{ij}$ & $|\mathbf{D}_{ij}|$ \\
   & & (i-j) & (meV) & (meV) & (meV) \\ \hline
   1L & \textbf{Intra} & & & \\
    & NN & All & -4.35 & 1.49 & 0.00 \\
      & NNN & All & -0.74 & 0.17 & 0.06\\ \hline
   2L-AA & \textbf{Intra} & & & \\
    & NN & All & -4.42 & 1.44 & 0.12 \\
   & NNN & All & -0.65 & 0.15 & 0.06 \\
    & \textbf{Inter} & & & \\
   & NN & All & -0.12 & 0.08 & 0.06 \\
   & NNN & All & -0.13 & 0.02 & 0.00 \\ \hline
   2L-AB & \textbf{Intra} & & & \\
   & NN & (1-2) & -4.62 & 1.45 & 0.13 \\
   & NN & (2-3) & -4.20 & 1.39 & 0.05 \\
   & NN & (2-5) & -4.64 & 1.50 & 0.11 \\
   & NNN$_{\mathrm{A}}$ & All & -0.59 & 0.15 & 0.05 \\
   & NNN$_{\mathrm{B}}$ & All & -0.65 & 0.14 & 0.06 \\
       & \textbf{Inter} & & & \\
   & NN$_{\mathrm{B}}$ & All & -0.31 & 0.11 & 0.00 \\
   & NNN & \footnote{\ Pairs: (7-11), (7-13), (7-15), (8-12), (8-14), (8-16)} & -0.33 & 0.05 & 0.01 \\
   & NNN & \footnote{\ Pairs: (7-10), (7-12), (7-14)} & -0.50 & 0.15 & 0.00 \\ \hline
   2L-AB' & \textbf{Intra} & & & \\
    & NN & (1-2) & -4.62 & 1.47 & 0.12 \\
   & NN & (2-3) & -4.20 & 1.39 & 0.05 \\
   & NN & (2-7) & -4.64 & 1.49 & 0.12 \\
   & NNN & (1-3), (4-8) & -0.68 & 0.16 & 0.04 \\
   & NNN & (1-7), (2-4) & -0.62 & 0.15 & 0.04 \\
   & NNN & (3-7), (2-8) & -0.62 & 0.15 & 0.05 \\
       & \textbf{Inter} & & & \\
   & NN & (7-20), (2-19) & -0.31 & 0.10 & 0.00 \\
   & NN & (1-19), (2-20) & -0.31 & 0.10 & 0.05 \\
   & NNN & (9-26), (10-29) & 0.16 & 0.06 & 0.00 \\
   & NNN & (9-21), (10-22) & 0.11 & 0.05 & 0.04 \\
   & 3NN & (9-20), (10-23) & 0.12 & 0.01 & 0.00 \\
   & 3NN & (9-23), (10-24) & 0.05 & 0.06 & 0.04 \\ 
   & 3NN & (9-25), (10-26) & 0.05 & 0.03 & 0.09 \\
   & 3NN & (9-28), (10-33) & 0.01 & 0.01 & 0.00 \\ \hline\hline
\end{tabular}
\end{table}

\newpage
\section{Artificial tuning of the magnonic dispersion}\label{supsec:tuning}
As discussed in the main text, we can `tune' the spin-wave dispersion by artificially changing the magnetic parameters. However, note that, changing one specific parameter while keeping all the other parameters constant is very artificial. Real tuning of a material, e.g. by strain or gating, will inevitably influence multiple interactions. Nonetheless, the results presented in this section sketch a good general picture of how the magnonic bandgap depends on the different magnetic parameters. 

\begin{figure}[!ht]
  \centering
  \subfloat[][]{\includegraphics[width=.33\textwidth]{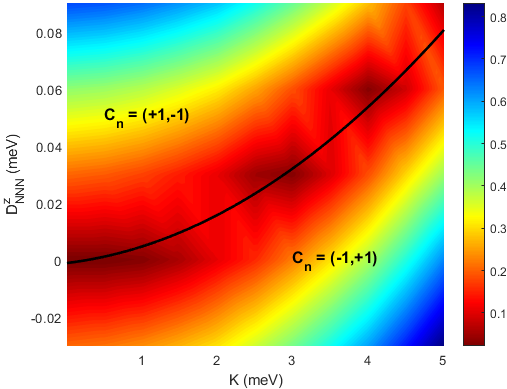}}
  \subfloat[][]{\includegraphics[width=.33\textwidth]{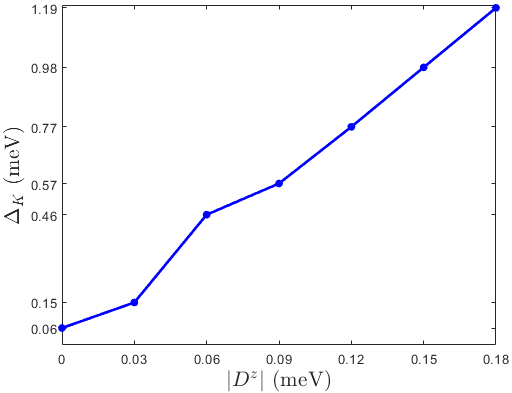}}
  \subfloat[][]{\includegraphics[width=.33\textwidth]{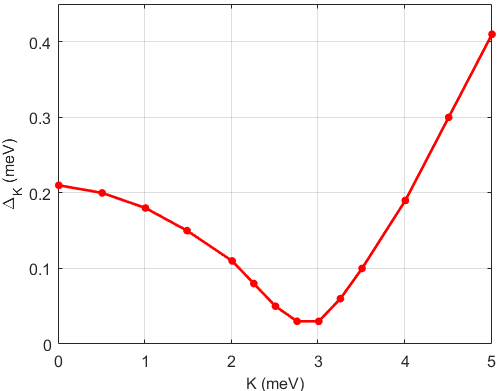}}\\
  \subfloat[][]{\includegraphics[width=.31\textwidth]{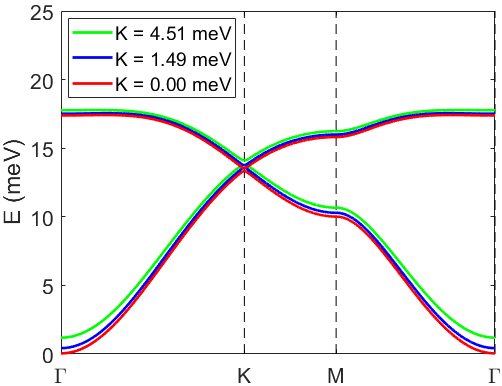}} \qquad
  \subfloat[][]{\includegraphics[width=.33\textwidth]{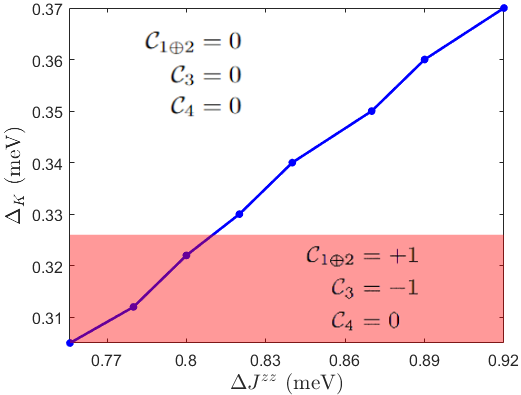}}
  \caption{\textbf{Artificial tuning of the magnonic dispersion by manually changing the magnetic parameters.} (a) Size of the magnonic bandgap plotted as a function of the NNN DMI and the NN Kitaev interaction. The black line marks the phase boundary between two topological phases with opposite Chern numbers for each band. (b) Size of the magnonic bandgap in monolayer CrI$_3$ plotted as a function of the NNN DMI (for $K_{\mathrm{NN}} = 1.49$ meV). (c) Size of the magnonic bandgap in monolayer CrI$_3$ plotted as a function of the NN Kitaev interaction (for $D^z_{\mathrm{NNN}} = 0.03$ meV). (d) Magnonic dispersion of monolayer CrI$_3$ for different values of the Kitaev constant. (c) Magnonic bandgap in an AB-stacked CrI$_3$ bilayer with FM interlayer ordering as a function of the out-of-plane exchange difference between the two sublattices. Corresponding (composite) Chern numbers are indicated for the topological trivial phase (white) and the topological non-trivial phase (red).}
  \label{fig:artificial_tuning}
\end{figure}

As shown in Figure \ref{fig:artificial_tuning}(a,b,c) the size of the magnonic bandgap in monolayer CrI$_3$ is tunable by changing the strength of the NNN DMI and the NN Kitaev constant. The phase diagram in Figure \ref{fig:artificial_tuning}(a) shows that, depending on the specific combination of DMI and Kitaev interaction magnitudes, we observe one of two topological phases with opposite Chern numbers for each band, which signifies that the propagation direction of the magnon edge modes is reversed. The overall trend that becomes apparent from this graph is in good agreement with some recent work that discusses the interplay between the DMI and the Kitaev interaction in honeycomb ferromagnets\cite{zhang2021}. Further, note that changing the Kitaev constant influences the shape of the dispersion in the whole Brillouin zone [Fig. \ref{fig:artificial_tuning}(d)], whereas the NNN DMI mainly effects the dispersion around the K-point and the K'-point. Similar bandgap tuning can be observed by artificially changing parameters in bilayer CrI$_3$. In Figure \ref{fig:artificial_tuning}(e), we show for an FM ordered AB-stacked bilayer that, by decreasing the out-of-plane exchange difference between sublattices $\Delta J^{zz} = |J^{zz}_\mathrm{A} - J^{zz}_\mathrm{B}|$, which is caused by the breaking of sublattice symmetry, we can also decrease the size of the magnonic bandgap and induce a topological phase transition from a topologically trivial to a topologically non-trivial phase. Likewise, by decreasing $\Delta J^{zz}$ in the AB'-stacking, a similar phase transition can be observed by opening/closing and size-tuning of the bandgap.

